# True Solid-State Electrical Conduction of Proteins shows them to be Efficient Transport Media

Sudipta Bera,[*,†] Ayelet Vilan,[§] Sourav Das,[†] Israel Pecht,[‡] David Ehre,[†] Mordechai Sheves,[*,†] and David Cahen[*,†]

[†] *Department of Molecular Chemistry and Materials Science, Weizmann Institute of Science, Rehovot 7610001, Israel*

[§] *Department of Chemical Research Support, Weizmann Institute of Sciences, Rehovot 7610001, Israel*

[‡] *Department of Immunology and Regenerative Biology, Weizmann Institute of Science, Rehovot 7610001, Israel*

*Corresponding Authors–email: sudipta.bera@weizmann.ac.il, mudi.sheves@weizmann.ac.il, david.cahen@weizmann.ac.il

ORCID

Sudipta Bera: 0000-0001-7894-9249
Ayelet Vilan: 0000-0001-5126-9315
Sourav Das: 0000-0002-4478-800X
Israel Pecht: 0000-0002-1883-9547
David Ehre: 0000-0002-5522-8059
Mordechai Sheves: 0000-0002-5048-8169
David Cahen: 0000-0001-8118-5446

**Abstract**

While solid-state protein junctions have shown efficient electron transport over lengths that surpass those of conventional organic semiconducting systems, interfacial /contact effects may obscure the intrinsic protein charge transport properties. Therefore, contact resistance ($R_C$) effects need to be quantified and then minimized, which poses a problem if 4-probe geometries cannot be used. Here we show how $R_C$ can be extracted quantitatively from the measured junction resistance ($R_P$) by using the extrapolated zero-length resistance ($R_{ZLR}$) and short-circuit resistance ($R_S$). We used AC (impedance spectroscopy) and DC measurements to examine charge transport in junctions of human serum albumin (HSA) and bacteriorhodopsin (bR) films with varying thicknesses. Three contact configurations, Si-Au, Au-EGaIn, and, in a micropore device (MpD), Au-Pd, were compared. While Si-Au and Au-EGaIn junctions exhibit substantial $R_C$ that we ascribe to interfacial oxides and electrostatic protein-electrode interactions, MpD effectively eliminates $R_C$, enabling measuring the intrinsic electron transport across HSA and bR films. The exponential length dependence of $R_P$ shows a transport decay constant ($\beta$) that varies with interfacial conditions, underscoring the role of contact engineering. By minimizing $R_C$, exceptionally low $\beta$ values (~0.7–1.1 nm$^{-1}$) are found, proving that, indeed, proteins can have outstanding charge transport efficiencies.

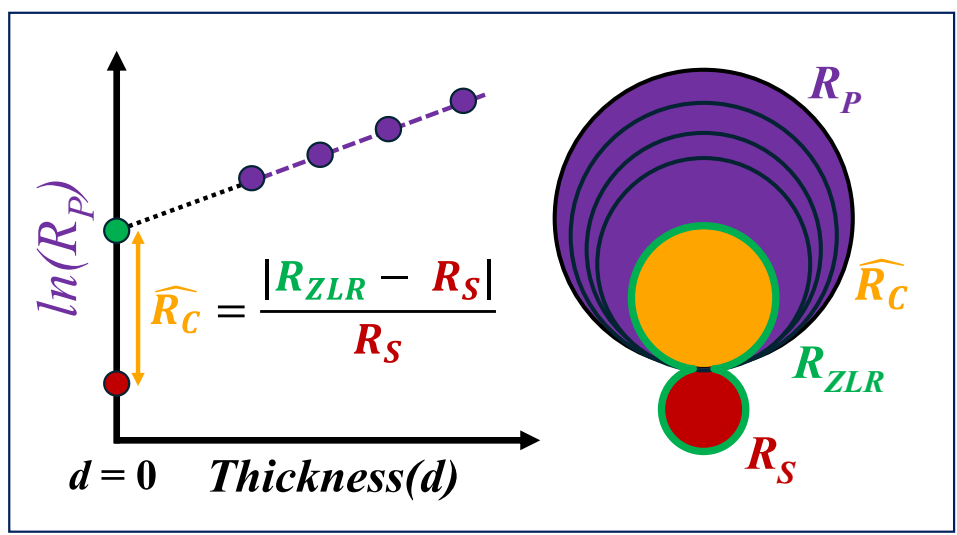

# 1. Introduction

Recent reports on electron transport across protein films, with transport distances spanning tens of nanometers,[1–5] suggest that proteins in solid-state junctions serve as a surprisingly efficient electron transport medium. The fact that these distances far exceed expectations for a medium with a significant saturated organic component, coupled with the lack of temperature dependence,[2,6] challenges our understanding of existing transport mechanisms. A theoretical cascade model, proposed by Papp and Vattay[7] predicts that a single contact dominates the transport via nm's-long proteins. In this model, the more resistive contact, rather than the protein itself, largely governs overall transport efficiency.[2] Futera et al.,[8] using density functional theory (DFT) and molecular dynamics showed that small (<3 nm) tetraheme protein junctions can function as one complete contact-protein-contact composite.

Testing a model for electron transport through proteins requires experimental determination of the intrinsic transport properties of the protein in the junctions. To that end, junctions should be such that the contribution of the contacts to the measured transport is minimal. This challenge is not unique to protein junctions; in molecular electronics, interfacial effects often overshadow intrinsic molecular properties defining characteristic of the field since its inception. In this regard, molecular junctions are not too different from other junctions, as demonstrated in textbooks on metal-metal contacts (see, e.g., ref.[9]).

A molecular junction can be conceptually divided into three key components: the two electrode-molecule contact interfaces and the molecular layer between them. Salomon et al.[10] demonstrated that electron transmission across molecules is primarily governed by the molecule-electrode interfaces and the intrinsic properties of the electrode materials. Indeed, as Hipps noted for DNA junctions, molecule-contact interfaces dictate conductivity, leading to reports of insulating, semiconducting, conducting, and even superconducting behavior.[11]

Achieving control over interfaces or contacts is challenging both conceptually and technically. How contacts are established can often have a more significant effect on molecular junction measurements than the molecules themselves. Even minor variations in a contact formation protocol can lead to substantial changes in device characteristics.[12] The nature of the contacts determines if any material mixing across the interface can occur or if only electronic charge carriers will cross. Here nature of contacts includes surface roughness (and, consequently, interface roughness), defect formation, activation or mitigation, and, in the case of soft materials, flexibility.[13,14] Therefore, to accurately assess the intrinsic charge transport capabilities of molecular junctions, contact effects should be first quantified and then minimized. This conclusion applies also to protein-based junctions.

In this study, we use the concept of contact resistance ($R_C$) to account for the effects of contacts and their interfaces with proteins. Typically, $R_C$ is mitigated using four probe measurements.[15] However, for single molecule or ultrathin film systems, implementing a four-probe setup would require electrodes spaced only a few nanometers apart, without any crosstalk—an approach that remains technically unfeasible. As a result, studies of molecular junctions have employed two different strategies for evaluating the different contributions to the junction resistance. One approach is the concept of zero-length resistance ($R_{ZLR}$), which is obtained by measuring the total resistance across junctions that are identical except for sample thickness (i.e., electrode-to-electrode separation), and exponentially extrapolating it to zero sample length.[16,17] $R_{ZLR}$ can be derived from length-dependent electron transport measurements of either DC or AC measurements. The other approach uses impedance measurements (Section S1 in SI) to separate the resistance at short-circuit, i.e., if the two electrodes are in direct contact, without a sample in between them, which is taken to give the circuit resistance ($R_S$), and the junction resistance ($R_P$, in parallel with a capacitor).[18] We propose here that the combination of these two approaches reveals the contribution of the contact to the junction resistance. Scheme 1 presents individual circuit element[2,19] details, including $R_C$ as part of $R_P$.

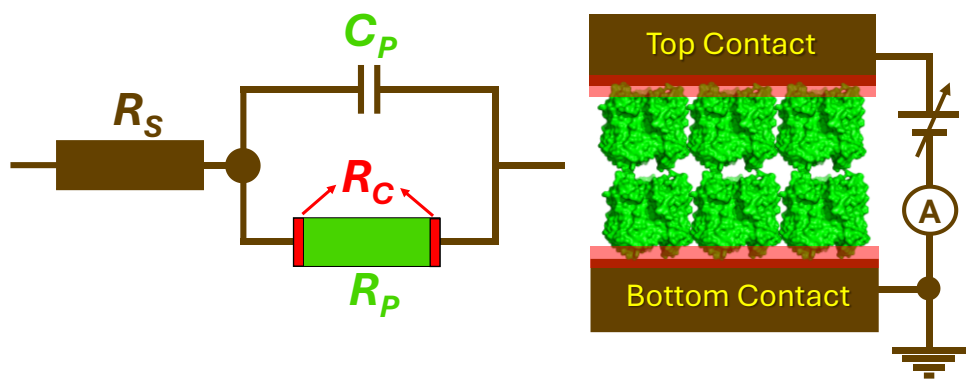

**Scheme 1**: Equivalent circuit of solid-state protein-based devices, irrespective of protein type and junction configurations; circuit resistance ($R_S$), protein resistance ($R_P$), and protein capacitance ($C_P$). Specific color coding represents different circuit components; green- protein part, dark brown- whole circuit with terminal leads and external wire connection without protein, and red-protein-electrode interfacial contact. In practice the value that is measured as $R_P$ includes $R_C$.

While formally the actual deviation between $R_{ZLR}$ and $R_S$ defines $R_C$, because $R_S$ will normally be very small compared to the other resistances in the circuit, $R_C$ will be dominated by $R_{ZLR}$, which makes comparing $R_C$ values between junction configurations problematic. To ensure scalability and meaningful comparison, here we reformulate $R_C$ as a dimensionless measure (represented as $\widehat{R_C}$) that reflects the difference between $R_{ZLR}$ and $R_S$ relative to $R_S$, as follows:

$$\widehat{R_C} = \frac{|R_{ZLR} - R_S|}{R_S} \quad (1)$$

Engelkes *et al.*[17] used the $R_{ZLR}$ directly as the $R_C$ for metal/insulator/metal junctions in a conductive atomic force microscopy (C-AFM) configuration. Notably, $R_{ZLR}$ varies with contact type;[16] it is strongly influenced by the metal electrode's work function,[17] emphasizing the difference between 'contact-limited' and 'contact-influenced' transport. The latter includes the dependence of transport rates on contact details, such as the density of electronic states and their coupling strength to the medium. 'Contact-limited' refers to cases where the injection of carriers across the interface limits the net transport. Admittedly, it is not trivial to separate between these two aspects. Within the above-suggested definition of 'contact resistance,' cases where $\widehat{R_C} \to 0$ implies that the net transport is not 'contact-limited.'

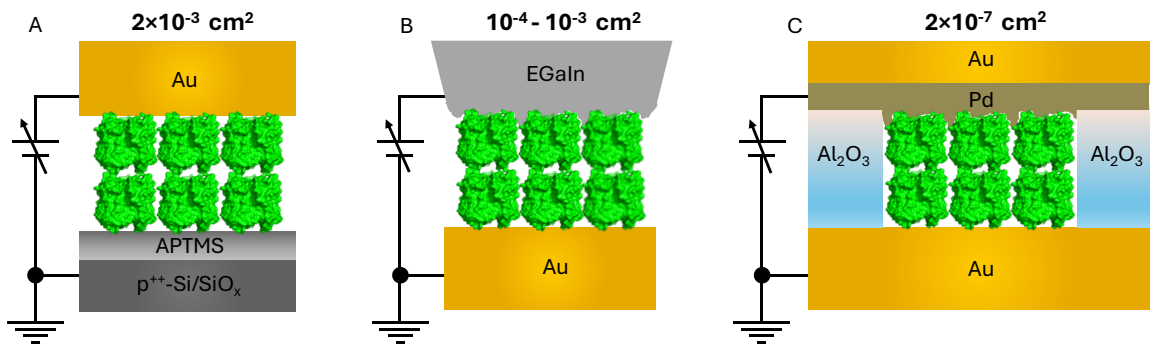

**Scheme 2**: Three different device configurations of both HSA and bR junctions with a cartoon representation of protein layers (A): *p$^{++}$-Si(SiO$_x$)/APTMS/protein/Au$^{pad}$* (Si-Au), (B): *Au/linker/protein/EGaIn$^{cone}$* (Au-EGaIn), and (C): *Au/ linker/protein /(Pd-Au)$^{MpD}$*. The respective geometric junction areas are indicated above the scheme of each device. Here HSA binds directly to the bottom Au-electrode without a linker layer (via exposed cysteine residues), whereas bR is immobilized on Au via a cysteamine linker in both the Au-EGaIn and MpD configurations.

Also, the (short) circuit resistance ($R_S$) can be evaluated by either AC or DC measurements. The AC impedance-fitted equivalent circuit yields $R_S$, which, in principle, should equal the resistance of the corresponding 'empty' junction, i.e., shorted but otherwise of identical geometry and configuration as the junctions used to measure the samples. However, in practice, this common-sense identity is not always observed, specifically for junctions with an EGaIn electrode where the impedance-derived $R_S$ often deviates by several orders of magnitude from the DC-shorted (DC-$R_S$) junction one.[20] The reason for this discrepancy could be limitations of the impedance technique applied to junctions with a semi-liquid (mechanically unstable) contact.

Here we present an approach to quantify the influence of $\widehat{R_C}$ by comparing both DC and AC transport measurements of systematically varying protein multilayers and testing its application on three distinct protein junction configurations (*see Scheme 2*):

$p^{++}$-Si(SiO$_x$)/linker/protein/Au$^{pad}$ , **Si-Au**;

Au/linker/protein/EGaIn$^{cone}$ , **Au-EGaIn**, and

Au/linker/protein/(Pd/Au)$^{MpD}$ in a micropore device, **MpD**.

The Si-Au and MpD junctions exhibit good reproducibility between AC and DC measured values (detailed methodology is given in SI Section S2), while for the Au-EGaIn junction, we use exclusively DC measurements due to significant mismatches and liquid-related impedance limitations (see Section 2.1.1 in the main text and Section S2.2.1 in SI).

To establish the validity of the $R_C$ method and genuine variation in protein transport properties, we investigated two distinct protein systems. The first is human serum albumin (HSA), a cofactor-free globular protein, chosen as a model system due to its advantages in junction preparation (discussed in SI Section S3). In parallel, we examined bacteriorhodopsin (bR), a retinal chromophore-embedded membrane protein. The bR system was recently reported to show unusual near-activation-less electron transport over 60 nm thick multi-layer protein samples in a Si-Au configuration.[2] Our key finding is that the MpD configuration effectively eliminates $R_C$ for the protein junctions, which enables getting at intrinsic protein transport characteristics. Additionally, we reanalyzed published studies ranging from small molecules to peptides, to test the validity of our approach. Overall, our results show the dominant role of interface electrostatics, while direct or linker-mediated protein-electrode contact, effective electrical contact, junction area, and electrode conductivity have a smaller influence on $R_C$.

## 2. Results and Discussion

In the SI, we give a comprehensive protocol for the preparation of the HSA and bR protein layers (see SI Sections S4 and S5) and for device fabrication (see experimental), including their surface and electrical characterization (see Section S6-S9).

### 2.1. Insights into Protein Circuit Elements

The fundamental circuit model of the dry protein junction is depicted in Scheme 1. A simplified equivalent circuit for the protein junction[2] was constructed based on impedance fitting, as elaborated in the SI (Section S1 and S2.2).

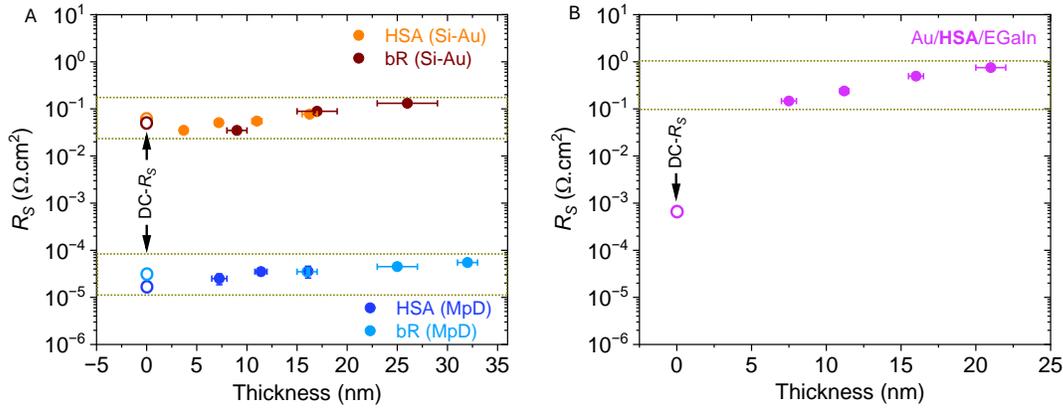

**Figure 1:** Impedance-derived, averaged $R_S$ values for different protein junctions (including both HSA and bR, with/without linker) are shown by color-filled circular dots, while the $DC$-$R_S$ values for the bare devices (shorted junction) are represented by hollow circular-color dots derived from $J$-$V$ slope (DC-measurement). (A) Different protein junctions with Si-Au and MpD device configurations are indicated in figure legends with the proper color. (B) for Au/HSA/EGaIn junctions.

### 2.1.1. Circuit Resistance ($R_S$)

In this study, $R_S$ is a critical parameter for evaluating contact resistance and ensuring the reliability of impedance measurements. At high frequencies (~1 MHz), the total junction resistance ($R_T$) is effectively reduced to $R_S$ because the AC current primarily flows through the capacitor, shorting $C_P$ in the parallel R-C circuit, while $R_p$ presents an infinitely resistive path, blocking the current (see Scheme 1). From DC measurements of shorted junctions (without a protein layer) the inverse slope of the J-V curve near zero voltage (see Figure S1) yields the DC-$R_S$ value (see SI Section S11). Notably, DC-$R_S$ closely aligns with the impedance-derived $R_S$ value (see Figure 1), confirming that $R_S$ reflects the properties of the terminal leads and external wire connections, independent of protein-electrode contact (see SI Section S1).

| Junctions | $R_S$ ($\Omega.cm^2$) | |
|---|---|---|
| | Impedance derived (Fitting) | DC measurement (shorted junction) |
| Si/SiO$_x$/APTMS/**HSA**/Au-pad | **5.0E-2** ± *2.0E-2* | **6.0E-2** ± *3.0E-2* |
| Si/SiO$_x$/APTMS/**bR**/Au-pad | **9.0E-2** ± *5.0E-2* | **3.0E-2** ± *2.0E-2* |
| Au/**HSA**/EGaIn | **4.0E-1** ± *3.0E-1* | **7.0E-4** ± *3.0E-4* |
| Au/ **HSA**/Pd-Au(MpD) | **3.0E-5** ± *2.0E-5* | **1.0E-5** ± *2.0E-5* |
| Au/cys/**bR**/Pd-Au(MpD) | **2.0E-5** ± *3.0E-5* | **3.0E-5** ± *3.0E-5* |

**Table 1**: Averaged $R_S$ values for various protein junctions, derived from impedance fitting, and the reciprocal slope of the *J-V* curves (see SI sec. S11) from *DC* measurements of electrically shorted junctions of the type indicated for each line.

To account for variations in junction area, $R_S$ is presented in an area-normalized format, as it is primarily influenced by the dimensions of the junction.[21] Here, we use all the resistance components in an area-normalized fashion based on the geometric area of the smallest contact of a junction. The extracted $R_S$ values follow the trend MpD < Au-EGaIn < Si-Au (see Table 1), a reasonable trend from the sum of the expected resistances of the two contacts. Given that one terminal electrode in each configuration is gold (Au), the observed differences in $R_S$ are attributed to the resistivity of the other electrode (Au/Pd for MpD < EGaIn < p$^{++}$-Si). For HSA and bR junctions, DC and AC $R_S$ values in the Si-Au and MpD configurations agree within a factor of 3 for a given protein or between proteins for a given contact configuration (Table 1, Figure 1).

In contrast, impedance-derived $R_S$ values for Au-EGaIn junctions deviate significantly (>10$^3$) from DC-$R_S$ (Figure 1, Table 1). A similar discrepancy (~10$^2$) was previously reported for EGaIn top contacts.[18,20] This suggests a common origin for the difference between AC and DC measurements on junctions with EGaIn contacts in our results and those from refs.[18,20], viz. the inability of the protein/molecular surface to establish a stable electrical interface with the semi-liquid EGaIn. Dynamic variations at the protein-EGaIn interface during frequency sweeps, coupled with insufficient adhesive forces between the protein/organic layer and EGaIn, likely contribute to this phenomenon. While DC values give the time- and space (contact

areas)-averaged results of a steady state that is established for the applied bias, reaching a steady state becomes problematic with AC modulation. Consequently, impedance-derived $R_S$ values for Au-EGaIn junctions become unreliable for comparative analysis (see more in SI Section S2.2.1). Therefore, throughout this study, DC-based measurements were exclusively used for these junctions.

### 2.1.2. Protein Resistance ($R_P$) and Interface Contributions

At low frequencies (<10 Hz), AC current traverses both resistive components in the equivalent circuit, while the capacitor ($C_P$) behaves as an open circuit, effectively blocking AC current. Consequently, the total junction resistance is measured as $R_T = R_S + R_P$ (a series combination) (see Scheme 1). $R_P$ values were extracted from the AC measurements via Nyquist plot analysis for each protein junction (see SI Section S1). In parallel, DC measurements provided an alternative estimate of $R_T$ (from the average J-V response obtained from Figure S2, described in SI Section S11), where the impedance-derived $R_P$ closely matched the DC-derived $R_T$ for both HSA and bR in Si-Au and MpD configurations (see Figure S3). This agreement arises because $R_T$ is predominantly governed by $R_P$, with only a negligible contribution from $R_S$ (i.e., $R_P \approx R_T$). However, as mentioned earlier (Section 2.1.1), due to experimental limitations in Au-EGaIn junctions, the total resistance (equivalent to $R_P$) can be reliably determined only through DC measurements.

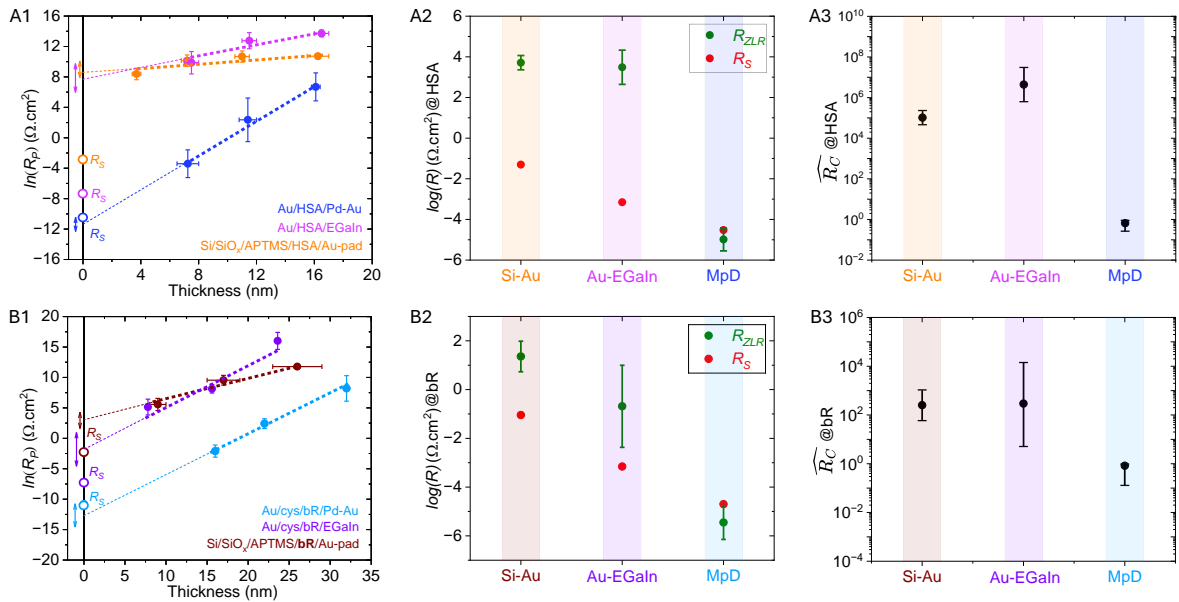

**Figure 2:** (**A1**), (**B1**) Plots of $ln(R_P)$ versus protein layer thickness, to estimate parameters for the three different types of protein junctions, each with (HSA or bR). The different devices are distinguished by their colors. The linear fits (dotted lines; $R^2 = 0.99$) yield the Y-axis intercept, representing $R_{ZLR}$. The colored double-headed arrows indicate

experimental errors (mean ± standard deviation), while the colored hollow circular dots at zero length correspond to $R_S$ of respective devices. Both X and Y error bars represent the mean ± standard deviation over 20 protein junctions per data set. The error bars for $R_S$ fall within the represented dots. (**A2**) and (**B2**) illustrate the relative contributions of $R_S$ and $R_{ZLR}$; (**A3**) and (**B3**) provide the estimated $\widehat{R_C}$ values (cf. Equation 1) for a specific junction and protein type.

Strikingly, for both HSA and bR junctions (Figure S3), switching from Si-Au to MpD configurations led to a decrease of several orders of magnitude in $R_P$, particularly for the thinner (8 and 16 nm) protein junctions (Figure 2). This finding indicates that $R_P$ is not solely dictated by the intrinsic properties of the protein layer. Instead, there are likely significant contact effects, arising primarily from the protein-electrode interface. Thus, $R_P$ encompasses contributions from both the bulk protein layer and the protein-electrode contacts. To mitigate interface effects, an extensive contact-engineering study was conducted, measuring electron transport (ETp) across HSA and bR junctions in each of the three distinct device configurations.

**Table 2A**

| Junctions | $R_{ZLR}$ ($\Omega \cdot cm^2$) | $\widehat{R_C}$ | $\beta$ (nm$^{-1}$) |
|---|---|---|---|
| Au/**HSA**/Pd-Au | 3.2E-6 – 3.4E-5 | 8.00E–01 | 1.10±0.05 |
| Au/**HSA**/EGaIn | 4.4E+2 – 2.1E+4 | 4.4E+06 | 0.35±0.10 |
| Si/SiO$_x$/APTMS/**HSA**/Au-pad | 2.3E+3 – 1.2E+4 | 1.00E+05 | 0.10±0.05 |

**Table 2B**

| Junctions | $R_{ZLR}$ ($\Omega \cdot cm^2$) | $\widehat{R_C}$ | $\beta$ (nm$^{-1}$) |
|---|---|---|---|
| Au/**bR**/Pd-Au | 8.1E-7 – 1.5E-5 | 9.60E-01 | 0.70±0.05 |
| Au/**bR**/EGaIn | 4.3E-3 – 1.0E+1 | 2.90E+02 | 0.70±0.10 |
| Si/SiO$_x$/APTMS/**bR**/Au-pad | 5.4E+0 – 9.6E+1 | 2.50E+02 | 0.30±0.05 |

**Table 2**: Parameters extracted for the different protein junctions from $lnR_P$ vs. thickness plots from Figure 2. Table 2A: HSA; Table 2B: bR. Here, the $\widehat{R_C}$ values are unitless as they refer to the relative deviation of $R_{ZLR}$ from $R_S$ for a specific junction and protein type.

## 2.2. Length Dependence of $R_P$ and Introduction of $R_{ZLR}$ and $\widehat{R_C}$

The resistance of protein-based junctions ($R_P$ or $R_T$) depends exponentially on protein layer thickness ($d$), following a general trend irrespective of experimental device configurations (Figure 2). Extrapolating to $d$ = 0 reveals a finite resistance at the y-axis intercept, termed the "zero-length resistance" ($R_{ZLR}$). Empirically, $R_P$ can be expressed as:

$$R_P = R_{ZLR} \cdot e^{\beta \cdot d} \qquad (2)$$

where $\beta$ is a distance decay constant (see Table 2), and $R_{ZLR}$ serves as the pre-exponential factor. In this study, $R_P$ values were derived from impedance fitting for protein-based Si-Au and MpD junctions, while for Au-EGaIn configurations, DC-derived total resistance ($R_T \approx R_P$, given $R_P \gg R_S$) was used. The estimated $R_{ZLR}$ values for different protein types and configurations are presented in Table 2.

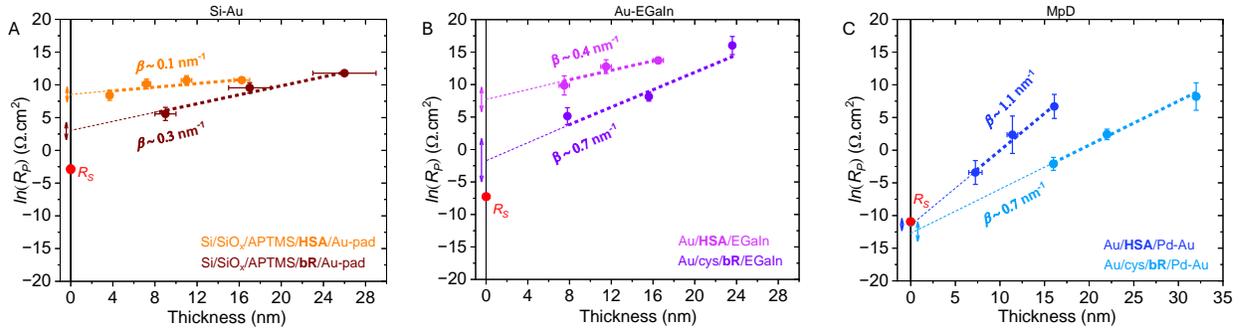

**Figure 3**: Comparing HSA and bR junctions in terms of the $ln(R_P)$ vs. thickness plots in all three device configurations, as indicated in the figure legends with distinct color schemes, the slope measures respective $\beta$ values (nm$^{-1}$). The red dot is the average circuit resistance (the error bars fall within the represented dots); a double-headed arrow indicates the estimated $R_{ZLR}$ (with experimental errors; mean ± standard deviation) by linear fitting. Separate device configurations are (**A**) Si-Au, (**B**) Au-EGaIn, and (**C**) MpD.

For HSA junctions, the average $R_{ZLR}$ is approximately two orders of magnitude higher than for bR junctions in Si-Au configurations, and four orders of magnitude higher in Au-EGaIn setups (Figure 3). However, MpD junctions exhibit significantly lower $R_{ZLR}$ values, with minimal variation between bR and HSA. $R_{ZLR}$ should predominantly reflect resistance contributions from leads and external circuitry, independent of the protein layer. Thus, it should ideally converge with $R_S$. However, deviations are evident across different configurations: while MpD junctions show near-convergence, Si-Au, and Au-EGaIn setups exhibit substantial discrepancies. As noted above, this difference is represented by the dimensionless form of contact resistance ($\widehat{R_C}$) (see Equation 1), i.e., the greater the difference $|R_{ZLR} – R_S|$, the higher is value of $\widehat{R_C}$. It is reasonable to attribute the variations in $\widehat{R_C}$ to chemical and electrical protein-electrode interactions, which will be affected by differences in protein surface charges. Notably, $\widehat{R_C}$, which can be due to the top or bottom interface, or both, is significant in Si-Au and Au-EGaIn junctions, where it constitutes a major part of $R_P$. In contrast, in MpD configurations $R_{ZLR} \approx R_S$, i.e., protein/contact interactions have minimal impact (see Figures 2 A3 and B3).

### 2.3. Origin of Contact Resistance ($\widehat{R_C}$) in Protein Junctions

The concept of "conductance" in molecular systems, including proteins, is often treated by analogy to bulk materials like copper or silicon. However, a fundamental question in ETp studies arises: are we truly measuring the intrinsic properties of the protein, or does contact resistance ($R_C$) play a dominant factor?

*2.3.1. Influence of $\widehat{R_C}$ on ETp*

Our experimental findings reveal that in ETp through both Si-Au and Au-EGaIn protein junctions, $R_C$ is substantial. In Si-Au configurations, the relative deviation of $R_{ZLR} \approx 10^5 R_S$ for HSA and $\approx 10^2 R_S$ for bR. In Au-EGaIn junctions $R_{ZLR} \approx 10^6 R_S$ for HSA, while for bR $R_{ZLR}$ remains as in Si-Au junctions, $\approx 10^2 R_S$ (Figure 2, Table 2). Within the proposed circuit model (Scheme 1), $R_C$ can be positioned before or after $R_P$, but impedance analysis cannot distinguish between the two separately. Consequently, in the presence of significant contact resistance, $R_P$ is strongly affected, leading to $R_C$-dominated transport behavior.

While we cannot pinpoint the dominant origins of $R_C$, interface-specific factors are likely contributors. In Si-Au junctions, a ~1 nm silicon oxide[2] layer at the bottom interface increases resistance, while in Au-EGaIn junctions, a ~1 nm Ga-oxide[20,22] layer—whose resistivity is ~$10^8$ times[23,24] higher than that of EGaIn—similarly influences $R_C$. Interestingly, the presence of an interleaved ~0.5 nm amine-terminated molecular protein-electrode linker for bR (see SI Section S4.1 and S5.3) does not significantly alter $R_C$ in the MpD configuration, The role of a mechanically placed top electrode in modulating $R_C$ remains uncertain and is discussed later in Section 2.4.1.

Notably, although both Si-Au and Au-EGaIn configurations are highly contact-dominated, they still distinguish between different protein types (Figures 2, 3). For HSA junctions we deduce an $\widehat{R_C}$ value that is $10^3$-$10^4$ times higher than for bR ones, likely due to the known differences in protein surface electrostatics. Protein Data Bank (PDB) analysis (see SI Section S12) reveals that the exposed HSA surface is predominantly composed of polar and charged residues with minimal cavity areas, contrasting with bR (Figure S4). This high interfacial charge density may facilitate charge redistribution during transport, further amplifying $\widehat{R_C}$.

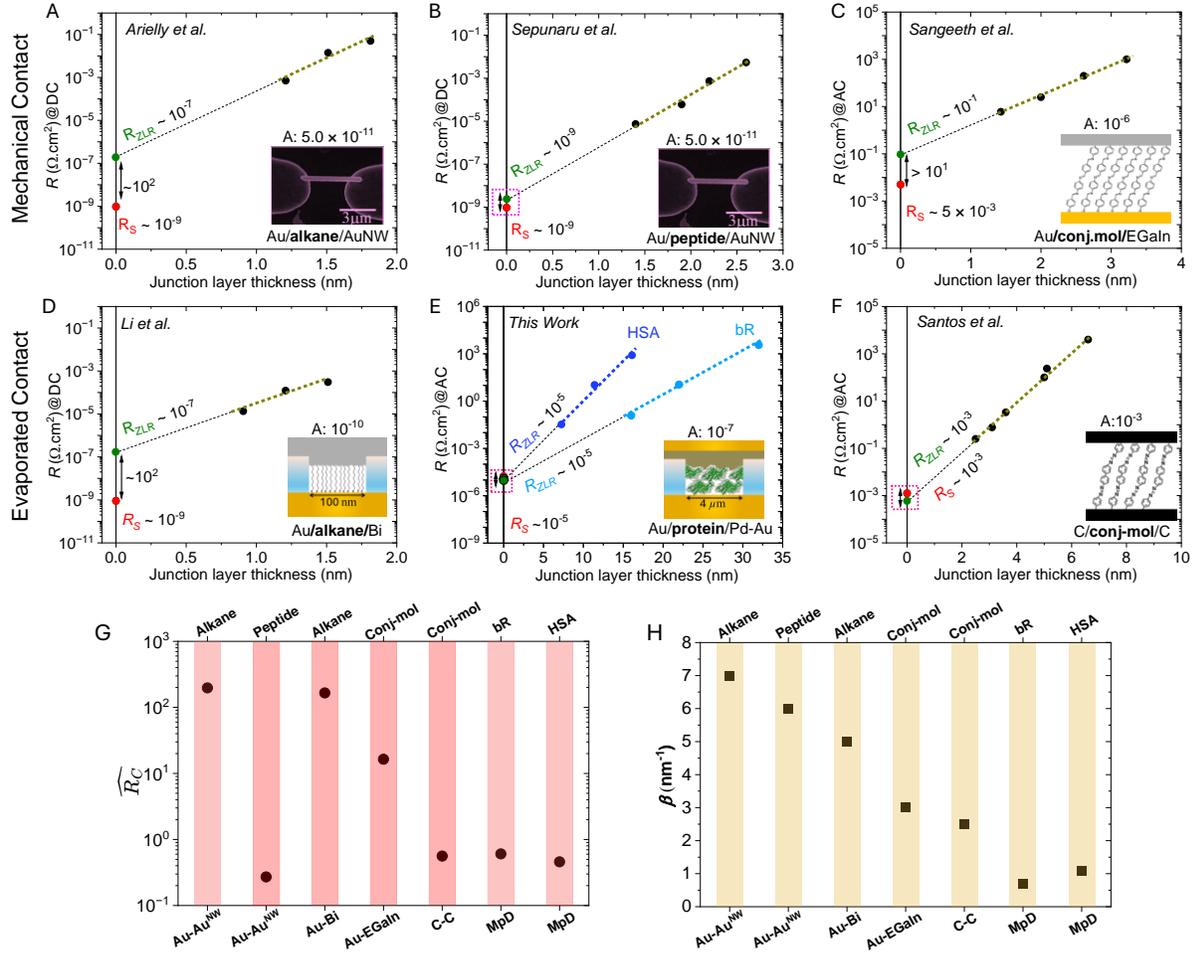

**Figure 4:** ln$R$ vs. $d$ plots for (A-D, F) various molecular junctions from the literature,[22,25–28] as indicated in the figure legends, to extract the area-normalized $R_{ZLR}$ values; the geometric junction area [cm$^2$] appears above the images/cartoons in each figure; (E) the MpD results for both bR and HSA configurations. The red and green dots correspond to $R_S$ and $R_{ZLR}$ values in [Ω.cm$^2$], respectively, along with the averaged estimated values. Specific device configurations appear in figure legends. Black arrows show | $R_{ZLR}$-$R_S$ |; a pink dotted box signifies the elimination of contact resistance. TOP: mechanical top contact configurations; BOTTOM: evaporated top contact configurations. Panels (A),[25] (B),[26] and (D)[27] show DC-based data; panels (C),[22] (E), and (F)[28] show AC-based data. NOTE: (A), (B), and (D) have the same ordinate; (C) and (F) have the same ordinate, while that for (E) is different as it spans 15 orders of magnitude, instead of 10 or 11. The estimated values of (G) $\widehat{R_C}$ (cf. Equation 1) and (H) current decay constant ($\beta$ nm$^{-1}$) from the slope of $lnR$ vs. thickness for seven different molecular-junction in published work (A-D, F) including two protein junctions from this work (E), each featuring a distinct sandwiched molecular/protein layer (indicated at the top along the X-axis). The terminal junction electrodes are specified at the bottom of the figure.

## 2.4. $\widehat{R_C}$ in Molecular Junctions

To validate our approach for extracting $\widehat{R_C}$ by disentangling it from $R_P$, we reanalyzed length-dependent resistance data from published literature.[22,25–28] Studies on protein junctions remain sparse, and, instead, we examined various small organic molecular to oligopeptide junctions across different contact configurations, as summarized in Figure 4. $\widehat{R_C}$ values differ significantly between Au/alkyl-thiol/Au(nanowire)[25] and Au/peptide/Au(nanowire)[26], to be denoted both as "Au-Au$^{NW}$", junctions. The alkane-based junctions exhibit an $\widehat{R_C}$ of ~100, whereas replacing alkane with a small peptide in the same Au-Au$^{NW}$ configuration eliminates $\widehat{R_C}$ ($\rightarrow$ 0) (See Figure 4G). Similarly, conjugated organic molecules (conj-mol) in Au/conj-mol/EGaIn[22]. "Au-EGaIn", junctions show notable $\widehat{R_C}$ (~20), while carbon/conj-mol/carbon[28](C-C) junctions display negligible $\widehat{R_C}$.

### 2.4.1. Influence of Junction Area on $\widehat{R_C}$

As significant variations in $\widehat{R_C}$ (spanning several orders of magnitude) arise across experimental configurations (Table 2) we can ask how far the differences in junction areas are responsible for these variations (up to $10^4$; Scheme 2), along with the nature of the top electrode contact. Note that 2 types (Si-Au, Au-EGaIn) are mechanically deposited in the ambient, while the other is vacuum-evaporated (MpD). contacts. As is known in the field of electrical (metal to metal) contacts, any of these differences can affect the electrical quality of the contacts.[9,29] Previous research established that molecular transport properties depend on the ratio of *electrically active* to *geometric* contact areas, which increases with decreasing geometrical area, approaching unity for STM studies.[10] Mukhopadhyay et al.[30] further reported significant inter-laboratory variations in junction current magnitudes (but not in the currents' temperature dependence), highlighting effective electrical contact as a critical factor. To assess the area effect on $\widehat{R_C}$, we conducted a comparative analysis of published data.

Despite an enormous (>$10^8$) junction area difference between the junctions reported and varying top contact types (mechanical vs. evaporated), Sepunaru et al.[26] and Santos et al.[28] showed negligible $\widehat{R_C}$. This suggests that $\widehat{R_C}$ is not dictated by the effective electrical contact area. Notably, in Au-Au$^{NW}$ configurations, replacing peptides[26] with alkyl thiols[25] introduces contact resistance, indicating that the electrode-molecule *interface* dominates over the contact area in determining $\widehat{R_C}$. Junctions with similar alkyl thiol molecules and comparable top contact areas yield nearly identical $\widehat{R_C}$ (~100), regardless of whether the top electrode is evaporated Bi (Au-Bi; $A_{geo}$ ~$10^{-10}$ cm$^2$, Li et al.)[27] or electrophoretically deposited Au (nanowire with $A_{geo}$ ~$10^{-11}$ cm$^2$, Arielly et al.).[25] Conversely, carbon/conj-mol/carbon[28] junctions with a substantially larger area ($1.3 \times 10^{-3}$ cm$^2$) exhibit negligible $\widehat{R_C}$, while Au/conj-mol/EGaIn[22] junctions with a $10^3$-fold smaller area ($10^{-6}$ cm$^2$) show a small but non-negligible $\widehat{R_C}$ (Figure 4, Sangeet et al.).[22] Therefore, the effective electrical contact area and its variations do not dominate $\widehat{R_C}$, whereas the interfacial characteristics do.

*2.4.2. Electrode Type Effects on $\widehat{R_C}$*

Interestingly, relatively high-resistance terminal electrodes ($R_S$ ~$10^{-3}$ Ω·cm$^2$) in carbon/conj-mol/carbon junctions do not influence $\widehat{R_C}$, whereas low-resistance terminal contacts ($R_S$ ~$10^{-10}$ Ω·cm$^2$) in Au/alkyl-thiol/Au(nanowire) junctions induce significant $\widehat{R_C}$. This underscores the dominant role of the molecule-electrode interface. In carbon/conj-mol/carbon junctions, the absence of an interface transport barrier due to possible direct C-C interactions likely minimizes $\widehat{R_C}$. In contrast, in Au/alkyl-thiol/Au(nanowire) junctions, interface dipoles are likely present between the terminal methyl group and the Au (possibly due to the spill-over electron density pillow effect),[31,32] contributing to contact resistance for the uncharged alkyl chains. However, charged molecules such as proteins and peptides can neutralize or reduce metal-molecule interface dipoles and electrostatic effects, thereby decreasing or eliminating $\widehat{R_C}$.

## 2.5. Intrinsic *β* Values of Protein Junctions

Experimentally, *β* can be estimated from the slope of the *ln*($R_P$) vs. *d* plots in Figure 2 for different junction configurations. Our results do not only show a way to eliminate contact resistance in protein junctions but also provide an accurate determination of the transport length decay constant (*β*) by taking into account $\widehat{R_C}$. With HSA *β* is an order of magnitude higher in the MpD configuration than in the Si-Au one, while *β* values for HSA Au-EGaIn junctions are in between these. For bR junctions *β* values appear less affected by contact configuration, with that in the MpD or EGaIn configurations roughly twice that in the Si-Au junctions (Table 2). Most likely, interface electrostatics control the charge injection of carriers into/extraction out of protein junctions. Then differences in *β* values reflect different interface charges.

Contact engineering plays a critical role in modulating charge transport through protein films. In HSA junctions within the Si-Au configuration, higher contact resistance weakens transport length dependence, resulting in lower *β* values, that inaccurately represent the protein's transport properties. Conversely, configurations with minimal contact resistance yield *β* values that more accurately reflect intrinsic protein transport characteristics. In the junctions dominated by contact resistance (Si-Au and Au-EGaIn), *β* values are initially higher for thinner films but decrease with increasing thickness (Figure S5). In contrast, junctions with negligible contact resistance (e.g., MpD) exhibit consistent *β* values across varying film widths.

The higher $\beta$ value observed for HSA ($\beta$ ~1.1 nm$^{-1}$) than for bR ($\beta$ ~0.7 nm$^{-1}$) MpD junctions suggests more efficient electron transport through bR than HSA layers. In the Si-Au configuration, this is the opposite, which we ascribe to contact resistance effects. Compared to small organic molecules and oligopeptides ($\beta$ ~3-7 nm$^{-1}$), proteins exhibit lower $\beta$ values (<1 nm$^{-1}$) even if, as is the case here, contact resistance is minimized (Figure 4H), reflecting their efficiency as electron transport media. The fact that this finding still stands now that contact resistance can be neglected, drives home the need to discover the mechanism(s) that make such efficiency possible.

## 3. Conclusions

This study highlights the critical role of contact/interface engineering in accurately assessing intrinsic electron transport (ETp) through proteins in protein-based junctions. By systematically eliminating contact resistance ($R_C$) in two-probe configurations, we demonstrate that interfacial electrostatics, interface dipoles, and insulating interfacial layers—rather than effective electrical contact area—primarily govern $R_C$. Our findings reveal that conventional contact-dominated configurations (e.g., Si-Au and Au-EGaIn) can obscure intrinsic transport properties due to interfacial oxides, whereas the micropore device (MpD) configuration effectively isolates and preserves the inherent charge transport characteristics of proteins, while having much better reproducibility than the Au$^{NW}$ contact approach. In contrast, the presence/absence of a linker or related protein-electrode bindings in an MpD configuration, do not need to lead to contact resistance. The consistent tunneling decay observed in $R_C$-free protein junctions proves that proteins can function as robust[33] charge transport media, comparable to or even surpassing traditional (organic) molecular junctions. Our approach offers a blueprint for designing molecular and bioelectronic interfaces with control over charge transport. Future research should integrate our $R_C$-free methodology with emerging bioelectronics to enable next-generation protein-based transistors, nanoscale energy harvesters, and molecular computing. As such, this work paves the way for scalable, tunable protein-based bioelectronics by linking fundamental charge transport studies to practical applications.

## 4. Experimental

### 4.1. Protein Layer Preparation

In this study, we systematically prepared Human Serum Albumin (HSA, Sigma) films, highlighting their advantages in detail (see SI Section S4 and S3). Similarly, bacteriorhodopsin (bR) layers were fabricated following a well-established protocol,[2] with additional procedural details provided in SI Section S5.

### 4.2. Protein-Based Device Fabrication

Three distinct vertical protein-based device architectures were fabricated for transport measurements, as illustrated in Scheme 2. The fabrication process for each configuration is briefly summarized as follows:

*4.2.1. Si-Au Configuration*

Ultra-thin gold pads (~2 × $10^{-3}$ cm$^2$) were manually placed onto protein layers deposited on linker-functionalized conductive silicon substrates, following previously established protocols.[2,34]

*4.2.2. Au-EGaIn Configuration*

A freshly prepared EGaIn (eutectic Indium-Gallium) cone was precisely positioned onto the protein layer using a micromanipulator, forming a large-area (~$10^{-4}$–$10^{-3}$ cm$^2$) top electrode at the protein-EGaIn interface. The setup is depicted in Figure S6, with additional details in SI Section S6.3.

*4.2.3. Micropore Device (MpD)*

A specialized protocol was developed for fabricating metal/protein/metal micropore devices (MpD),[33] designed to produce non-shorted, transport-active protein junctions. Recently, we demonstrated that in the MpD configuration, the underlying protein layer retained its functional activity despite the presence of the evaporated top electrode.[33] A photolithographically defined bottom Au electrode (~2 × $10^{-7}$ cm$^2$) was coated with protein layers, followed by successive E-beam evaporation of Pd and Au to form the top electrode. Further details are provided in SI Section S6.2.

**4.3. Electrical Characterization**

Electrical transport characteristics were investigated using both direct current (DC) measurements and alternating current (AC)-controlled impedance spectroscopy in a two-probe configuration. A comprehensive description of the measurement protocols is available in SI Section S2.

**Supporting Information**

The supporting information provides a comprehensive overview of the experimental methodologies and analyses used in this study. It includes impedance-derived equivalent circuits for protein devices (S1) and details of electrical measurements, covering both DC measurements (S2.1) and impedance measurements (S2.2), including limitations for EGaIn junctions (S2.2.1). The structural and preparation aspects of HSA (S3, S4) and bR thin films (S5) are outlined, including substrate cleaning, linker deposition, and multilayer formation. The top electrode deposition procedures (S6) for Si-Au junctions (S6.1), MpD devices (S6.2), and EGaIn-based electrodes (S6.3) are also described. The quality and thickness assessments of HSA/bR films (S7), along with impedance-based device evaluations (S8) and inverse capacitance-thickness relationships (S9), provide insights into protein layer characterization. Techniques for monitoring protein thickness (S10) via ellipsometry (S10.1) and AFM scratching (S10.2), as well as junction resistance analysis from J-V slopes (S11), are detailed. Finally, protein structural analysis (S12) offers insights into molecular features influencing charge transport.

**Author Contributions**

S.B. and D.C. conceptualized and designed the experiments in collaboration with D.E. and M.S. Experimental validation was carried out by D.E., S.B., D.C., A.V., and M.S. S.B. was responsible for the design, fabrication, optimization, and characterization of HSA-based devices and contributed to bR-based device development. S.B. conducted all electrical measurements (DC and impedance) on Si-Au and MpD junctions and performed data analysis for all protein-based devices. Additionally, S.B. developed the complete setup for EGaIn-based top electrode measurements and carried out HSA-based Au-EGaIn junction measurements. S.D. extracted and purified bR, prepared bR thin films, and conducted DC electrical measurements on bR-based Au-EGaIn junctions. Manuscript writing was contributed to by S.B., D.C., M.S., A.V., and I.P., with all authors actively participating in discussions and revisions.

**Acknowledgments**

D.C. and M.S. acknowledge financial support from the Deutsche Forschungsgemeinschaft (DFG) under the Middle East Collaboration Grant TO266/10-1. M.S. further extends appreciation to the Kimmelman Center for Biomolecular Structure and Assembly for additional support. We sincerely thank Eran Mishuk, Sigal Keshet, and Leonid Tunik for their valuable discussions on MpD device fabrication and optimization. We would like to acknowledge Sharon Garusi for providing engineering support for the development of custom-built instrumentation in the cleanroom. We gratefully acknowledge Sergey Khodorov for his contributions in developing an extended electrical setup and instrumentation. We also extend our appreciation to the dedicated Weizmann cleanroom facility (ISO 5-7), Weizmann physics core facilities, and Nisim Kanffo for assistance with shadow mask preparation. M.S. holds the Katzir-Makineni Chair in Chemistry.

# Supporting Information

## for

## True Solid-State Electrical Conduction of Proteins shows them to be Efficient Transport Media

Sudipta Bera,[*,†] Ayelet Vilan,[§] Sourav Das,[†] Israel Pecht,[‡] David Ehre,[†] Mordechai Sheves,[*,†] and David Cahen[*,†]

[†] *Department of Molecular Chemistry and Materials Science, Weizmann Institute of Science, Rehovot 7610001, Israel*

[§] *Department of Chemical Research Support, Weizmann Institute of Sciences, Rehovot 7610001, Israel*

[‡] *Department of Immunology and Regenerative Biology, Weizmann Institute of Science, Rehovot 7610001, Israel*

*Corresponding Authors–email: *sudipta.bera@weizmann.ac.il*, *mudi.sheves@weizmann.ac.il*, *david.cahen@weizmann.ac.il*

**Supporting Information Index**

## S1. Impedance-Derived Equivalent Circuits for Protein Devices

A common way for presenting impedance response is the Nyquist plot, which displays the imaginary component ($-Z_{im}$) versus the real component ($Z_{re}$) of impedance across a range of AC frequencies. For all three device configurations, the Nyquist plots of the protein junctions exhibit near-perfect semicircles, indicative of a single dielectric relaxation in the protein layers, as demonstrated for both HSA and bR junctions (Figure S7). Similar semicircular responses were observed in recent studies of bR junctions in Si-Au configurations.[1] These impedance characteristics can be modeled using a simple equivalent circuit: a parallel resistor-capacitor ($R$–$C$) element in series with a resistance, $R_S$ (see Scheme 1 in the main text). Here, $R_P$ represents the "protein resistance," influenced by interfacial contact, protein type, and layer thickness. $C_P$ denotes the "protein junction capacitance," which depends on the protein's dielectric constant, electrode separation, and junction area. $R_S$ accounts for the "circuit resistance" of the entire setup, including contributions from terminal electrodes but excluding effects from the protein or contact interface.

## S2. Electrical Measurements

In this work, we utilized both DC and impedance (AC) measurements to characterize various experimental protein junctions in a two-probe configuration. Si-Au and MpD-based protein devices were measured in a probe station (*LakeShore TTPX*) under a vacuum range of $10^{-3}$ to $10^{-5}$ mbar at room temperature (293 ± 2 K). Due to practical constraints, the Au-EGaIn junctions were measured under ambient conditions (< 40 % RH), as detailed in SI Section S2.2.1.

### S2.1. Direct Current (DC) Measurements

DC measurements were carried out by varying bias voltages using a sub-fA source meter (*Keithley 6430*) with an applied bias voltage ranging from ±0.1 V (1 mV step value) to ±1.0 V (10 mV step value) to the top electrode, while the bottom electrode was grounded. We executed the voltage sweep in a proper voltage loop starting from a zero voltage move to positive voltage maximum, then to the negative bias end through zero bias, and finally back to 0 V, using dedicated LabView-based programmable software with a scan rate 1mV/20 mS.

### S2.2. Impedance Measurements

Impedance spectroscopy (IS) measurements of the protein junctions were carried out at room temperature using a *Zurich Instruments MFIA* impedance analyzer. An AC bias with an amplitude ranging from 10 to 50 mV was applied with 0 V direct bias, and the frequency was swept from 1 Hz to 1 MHz at a rate of 15–20 data points per decade. The impedance data were analyzed in the form of Nyquist plots and phase vs.

frequency (a part of the Bode plot). Equivalent circuit elements were constructed from experimental data fitting using Z-view 4 software (*Scribner Associates*).

S2.2.1. *Limited Impedance Response Accusation for the EGaIn Junctions*

The impedance signal was noisy for the thicker protein layers (and there has not yet been any report of IS experiments with EGaIn contact with protein layer, thicker than 10 nm), and only the bilayer junction of HSA yielded reasonable signals (see Figure S7C). For the EGaIn junction, stable impedance data could be recorded only at a high frequency of ~1MHz, which approximately leads to the $R_S$ of the junctions, and we also extracted the junction $C_P$ values directly from the value of parallel capacitors in equivalent circuit module of MFIA connected LabOne software setup for different HSA junctions. We haven't tried this on bR junctions. The obtained large deviation between $R_S$ and DC-$R_S$, we restricted our measurements for EGaIn junctions for both the proteins, where we mainly focused on DC measurements.

## S3. Structural Advantages of HSA

Human serum albumin (HSA) possesses several features that make it highly suitable for transport studies. As a polypeptide-only protein, it readily forms mono- and multilayer junctions, aided by its numerous surface-exposed functional groups—such as cysteine, aspartic acid, glutamic acid, and lysine—which facilitate diverse substrate-protein linkages and inter-protein coupling (Figure S8). Additionally, HSA exhibits a unique binding affinity for various molecules, including ions, retinoic acid, and porphyrins, expanding its potential for future studies and biosensing applications.[2]

HSA is highly water-soluble and has an isoelectric point around pH 5, resulting in a net negative surface charge within the experimental pH range of 5.5–7. It displays a sharp UV-Vis absorption peak at approximately 280 nm (Figure S9). This negative charge supports strong immobilization on positively charged surfaces, such as amine-terminated linkers (e.g., cysteamine or APTMS). The abundance and uniform distribution of surface-active amino acid residues (Figure S8C–F) enable a variety of surface chemical reactions. Surface-exposed cysteine residues offer a notable advantage, allowing direct covalent binding to coinage and noble metals (Au, Ag, Cu) via metal–sulfur (M–S) bonds, eliminating the need for intermediate linkers. Furthermore, the presence of both acidic and basic amino acids supports efficient EDC-mediated coupling reactions, forming inter-protein amide bonds crucial for multilayer assembly. The asymmetric molecular structure of HSA (PDB: 1BM0)[3] and the even distribution of its active residues allow for multiple immobilization orientations. Crystal structure data indicates that HSA's longest dimensions span ~7.8 nm along the x- and y-axes (Figure S8A) and ~3.8 nm along the z-axis (Figure S8B). Interestingly, the observed layer thickness, regardless of substrate or the number of layers, was consistently ~4 nm, as measured by ellipsometry and AFM, except for the first layer on Au (~3 nm). This suggests a preferred orientation where the x–y plane serves as the thermodynamically favorable binding surface. Such orientation enables optimal interaction with both the substrate and adjacent protein molecules, supporting efficient multilayer formation.

**S4. HSA Thin Film Preparation**

**S4.1. HSA-SAM Preparation**

Self-assembled monolayers (SAMs) of HSA were prepared on three different substrates: $p^{++}$-Si/SiO$_x$, gold (Si/Au), and MpD with a bottom Au electrode. On the Si–SiO$_x$ substrate, HSA was electrostatically immobilized onto a positively charged, amine-terminated 3-aminopropyltrimethoxysilane (APTMS, Sigma) linker layer, which was attached to a <1 nm regrown silicon oxide layer on a highly doped (>10$^{20}$ cm$^{-3}$) Si wafer. The detailed procedures for preparing the substrate and applying the APTMS layer are described in recent work.[1] The APTMS-coated $p^{++}$-Si/SiO$_x$ substrate was incubated in a standard HSA solution (2 mg/mL in 20 mM phosphate buffer with 150 mM NaCl, pH 5.5) for 6 hours. After incubation, the electrostatically bound protein layer (Figure S10) was thoroughly rinsed with Milli-Q water and dried under nitrogen gas.

The same immobilization method was employed for HSA on both the Si/Au and MpD substrates, but without the use of linker molecules, as both possess Au-coated surfaces. First, the substrates were cleaned by sequential sonication in acetone, isopropanol (IPA), and Milli-Q water for 3 minutes each. This was followed by a 10–15 second treatment with hot base piranha solution (H$_2$O: H$_2$O$_2$: NH$_3$ = 5: 1: 1 v/v) at 80 °C to activate the Au surface for protein immobilization. After treatment, the substrates were thoroughly rinsed with Milli-Q water and dried with nitrogen gas. The activated surfaces were immediately incubated in the standard HSA solution (as described above) for 12 hours. Following incubation, the HSA-modified substrates were rinsed with Milli-Q water and dried under a flow of N$_2$ gas.

**S4.2. Preparation of HSA Multilayers**

HSA multilayers were prepared on various substrates using EDC coupling, as described in prior studies.[1] The procedure began with the activation of the HSA monolayer by incubating it in a freshly prepared EDC solution (20 mg/mL in 20 mM phosphate buffer containing 150 mM NaCl, pH 5.5) for 30 minutes. Following incubation, the substrate was rinsed with phosphate buffer to remove any unreacted EDC from the protein surface. Care was taken to remove the excess buffer from the substrate's bottom side while ensuring that the EDC-treated protein surface remained slightly hydrated with trace amounts of the buffer. Complete drying of the EDC-activated surface was avoided, as this would result in the decomposition of activated carboxylate groups back to their unreactive carboxylate form. The buffer-wetted, EDC-activated protein surface was immediately treated with a standard HSA solution (2 mg/mL in 20 mM phosphate buffer with 150 mM NaCl, pH 5.5) and incubated for 6 hours. Following incubation, the substrate was rinsed with Milli-Q water and dried with nitrogen gas. The EDC activation and HSA incubation steps were repeated as needed to achieve the desired number of HSA layers. During the treatment with HSA solution, the activated carboxylate groups on the (immobilized) protein surface reacted with the amine groups of lysine residues exposed on adsorbate HSA molecules, forming covalent inter-protein amide linkages (see Figure S10). This procedure was highly reproducible, allowing precise control over HSA multilayer growth and thickness.

### S5. bR Thin Film Preparation

In parallel, we conducted a study employing bacteriorhodopsin (bR) treated with octylthioglucoside (OTG), as described in detail in our preceding publications.[1,4] The immobilization of bacteriorhodopsin (bR) was carried out in a bilayer fashion.[1] Protocols for preparing single bilayers and multilayers on linker-coated $p^{++}$-Si/SiO$_x$ and Au substrates are provided in ref. [1]. For bR multilayers in MpD, a slightly modified procedure was used as follows:

### S5.1. Substrate Cleaning

MpD substrates were sonicated sequentially in acetone, isopropanol (IPA), and Milli-Q water for 3 minutes each. This was followed by a 20-25 sec treatment with hot base piranha (H$_2$O: H$_2$O$_2$: NH$_3$ =5:1:1 by v/v) solution (80 °C) to activate the surface. The substrates were then thoroughly rinsed with Milli-Q water and dried using nitrogen gas.

### S5.2. Deposition of Linker Layer

The cleaned MpD substrates were incubated overnight in a solution of cysteamine hydrochloride (cys) (4 mg/mL) in a pH 7.4 phosphate buffer medium. Following the incubation process, the cys-treated MpD was extensively rinsed with Milli-Q water, followed by 30-sec bath sonication to remove unbound cysteamine, and dried with nitrogen gas.

### S5.3. bR Bilayer(s) Preparation

The cys-modified MpD substrates were then incubated in a 4 $\mu$M solution of octylthioglucoside (OTG)-solubilized bR in 10 mM phosphate buffer (PB) containing 0.1 M ammonium sulfate ((NH$_4$)$_2$SO$_4$) at pH 6.4 for overnight. After incubation,[1] the bR-modified MpD was rinsed with Milli-Q water and dried with nitrogen gas. This procedure resulted in the formation of a single bR bilayer on MpD. For the preparation of bR multilayers, the protocol described in ref. [1], was followed.

### S6. Procedure for the Top Electrode Deposition

The protein (HSA or bR) modified bottom electrode was subjected to three different top electrode depositions as follows.

### S6.1. Si-Au Junctions

For the Si-Au junctions, the top Au-pads were deposited on the protein layer immobilized on p$^{++}$-Si/SiO$_x$/APTMS substrates as described in a recent work.[1]

### S6.2. MpD Devices and Probe Connections

The fabrication of micropore device (MpD) structures was carried out as detailed elsewhere.[5] Each device contained 16–56 micro-junctions. A top view of the entire MpD chip and a cross-sectional schematic of a single MpD junction are presented in **Figure S11**.

The bottom electrode of each device, consisting of an exposed Au surface, was electrically isolated. Protein multilayers were selectively deposited onto the micropore areas (~10$^{-7}$ cm²) of the Au surface, while the surrounding regions were insulated with alumina (see details in ref.[5]). The top electrode was formed by semi-indirect E-beam deposition of a 35 nm Pd layer, followed by a 25 nm Au protective coating, using an advanced evaporator system (Angstrom Evaporator), as detailed in ref.[5]. This top electrode served as the common (central) contact for all micro-junctions, to which the bias voltage was applied, while the bottom electrode was grounded (Figure S11A).

### S6.3. EGaIn-Cone-Based Top Electrode Landing and Measurement Setup

HSA and bR protein layers were prepared on Au substrates as described in Sections S4 and S5. A home-built setup was used to prepare Au-EGaIn junctions. Freshly prepared EGaIn (from Sigma) micro-cones with a typical tip diameter of ~10$^{-4}$ cm² were formed using a conductive metallic cylindrical probe attached to a micromanipulator. The EGaIn micro-cone landed softly on the protein surface (under live optical microscope monitoring) to form the top electrode (see **Figure S5**).

The experimental setup was mounted on a vibration-isolated table equipped with a vertical microscope for visualizing the protein-EGaIn contact and precisely measuring the contact area. The EGaIn top contact was connected either to an AC impedance analyzer (Zurich Instruments MFIA) or to a DC source meter (Agilent B2911A) with fA-level noise sensitivity. The substrate-attached protein layer was properly grounded during all measurements; for experimental parameters, see details in Section S2.1.

## S7. Quality and Thickness of HSA/bR Films

The thickness of the protein layers was primarily monitored by ellipsometry (see Table S2). For the Si-Au and Au-EGaIn junctions, ellipsometry measurements were performed directly on the protein-coated bottom electrodes prior to top electrode deposition. Due to the limited size of MpD junctions, direct ellipsometric characterization of the active area was not feasible. Instead, protein layer thicknesses for these junctions were measured on simultaneously prepared protein films on Au substrates—identical to those used in micropore fabrication. Protein layer growth was found to be uniform across different substrates, with a thickness of ~4 nm per layer for HSA and ~8–9 nm per bilayer for bR.

Up to four successive HSA layers were studied. For the Si-Au junctions, transport behavior was measured for mono-, bi-, tri-, and tetra-layered HSA films. However, for junctions on gold substrates (Au-EGaIn and MpD), only bi-, tri-, and tetra-layer junctions were characterized. HSA monolayers on gold substrates resulted in electrically shorted junctions, which we attribute to nanometer-scale inter-protein voids.[6]

For bR, junctions with one, two, and three bilayers were studied in both the Si-Au and Au-EGaIn configurations. As with HSA, all single bR bilayer junctions using the micropore configuration were shorted. Consequently, double-, triple-, and quadruple-bilayer junctions were analyzed to obtain three reliable data points for length-dependent studies.

The quality of the HSA layers was assessed using tapping-mode AFM imaging (Nanoscope V Multimode AFM setup) under ambient conditions, as described previously.[1,7] The quality and AFM characterization of bR bilayers have also been reported earlier.[1,4] AFM images revealed dense and uniform coverage of protein molecules across the substrates, with comparable RMS roughness on both $p^{++}$-Si/SiO$_x$/APTMS and Au surfaces, including within the micropores (Figure S12). To further validate thickness measurements, AFM scratching experiments were performed on HSA bilayers on Au, HSA trilayers on MpD, and HSA tetralayers on APTMS-coated $p^{++}$-Si/SiO$_x$ (Figure S13). The protein thickness values obtained from these experiments were in close agreement with ellipsometry data (see Table S2) for both HSA and bR. AFM-scratching-based layer thickness measurements of various bR layers were studied in detail in our recent work.[1,5]

Additionally, PM-IRRAS (polarization modulation-infrared reflection-absorption spectroscopy) of HSA trilayers revealed characteristic amide I and amide II bands (Figure S9A), consistent with known protein spectra such as those of bR,[1] streptavidin,[7] azurin,[8] and others.[9] Detailed procedures for these analyses are provided in our previous work.[7]

## S8. Quality Assessment of Protein-Based Devices via Impedance Analysis

Impedance-derived phase plots provide valuable insight into the formation of high-quality protein junctions. As discussed in Section 2.2.1 of the main text, under an applied AC field, the high-frequency regime (>10 kHz) predominantly drives AC current through the capacitive ($C_P$) component of protein junctions (see Scheme 1 in the main text). Consequently, a characteristic AC phase lag approaching 90° is expected due to the partial capacitive nature of the junction, such signature was clearly observed in the experimental protein junctions (Figure S14).

Under these conditions, protein junctions behave equivalent to a capacitor, where the sandwiched protein layers act as a dielectric medium that electrically isolates the two terminal electrodes. This configuration gives direct evidence for the prevention of filamentous growth or direct conduction (shorted) between the terminal electrodes, which is particularly crucial for MpD configuration. As a result, impedance-derived phase plots are valuable tools for identifying transport-active junctions while excluding those that exhibit shorting or partial shorting, as discussed in our current work.[5]

## S9. Inverse Capacitance-HSA/bR Thickness Relation

The capacitance of a parallel plate capacitor depends inversely on the dielectric separation between the plates (in our case, the protein layer thickness). Therefore, a plot of capacitance ($C_P$) versus inverse film thickness ($1/d$) is expected to be linear,[1] as shown in Figure S15. The linearity is maintained nicely for Si-Au junctions of HSA layers (shown here) and bR layers (see ref. [1]). However, MpD junctions (for both HSA and bR) do not exhibit significant $C_P$ variation with $1/d$ (Figure S15C, S15D), which can be attributed to the alumina ($Al_2O_3$) insulator defining the geometry of the micropore devices (see in SI, Section S6.2, and Figure S11).[5]

A single MpD chip integrates 16 to 56 devices, each featuring a top (Pd/Au-evaporated) contact area of approximately 10,000 $\mu m^2$ (see Figure S11). Over 99% of the device area is composed of highly insulating ALD-deposited alumina ($Al_2O_3$, ~20 nm thick), which separates the top and bottom metal electrodes. A small central micropore (~0.2% of the total area), measuring approximately 20 $\mu m^2$ with a depth of ~20 nm, serves as the active site.[5] The build-up protein layers (HSA or bR) specifically within this micropore cover only a negligible fraction of the total device area. Alumina, with a dielectric constant of ~10,[10] has a significantly higher permittivity than the protein layers (~3–4).[11,12] Consequently, the overall device capacitance is dominated by the alumina, and variations in the protein layer thickness within the micropore have a minimal impact on the measured capacitance. This behavior was observed consistently in both HSA and bR junctions within the MpD configuration. In contrast, the impedance-derived capacitance of Si-Au and Au-EGaIn junctions is closely correlated with protein layer thickness, as the protein spans the entire device area in these configurations. For these junctions, the estimated dielectric constant was ~2, relatively low, possibly due to the absence of cofactors in HSA, and lower than what was observed in bR-based junctions (~5).[1]

## S10. Monitoring Protein Layer Thickness

The study on the length dependence of ETp necessitates precise control over the layer thickness in the experimental protein junctions. As detailed above (see Section S7), we monitored protein layer thickness using direct methods for Si-Au and Au-EGaIn junctions and indirect methods for MpD, employing ellipsometry and occasionally AFM scratching experiments. The detailed experimental methodologies have been previously described.[1,7]

### S10.1. Ellipsometry

The protein layer thickness was determined using the Cauchy model by fitting the psi-delta data over the wavelength range of 350–1000 nm.[1,7] This analysis was conducted with the software of the Woollam M2000 V ellipsometer.

### S10.2. AFM Scratching

High-force AFM contact mode imaging (applied force: 160–200 nN)[1,6,7,13] was used to scratch protein surfaces. Scratching experiments were performed on bilayer, trilayer, and tetralayer HSA deposited on Au substrates, MpD, and Si/APTMS, respectively. *Gwyddion 2.63* software was used for AFM image processing, including roughness calculation and image presentation.

## S11. Junction Resistance from *J-V* Slope

The junction current was quantified in terms of current density, normalized by the geometric contact area of the protein junctions as described in the main text. The reciprocal of the averaged *J-V* curve (within the linear portion near 0V applied bias) for various protein layers provided the area-normalized junction resistance (in $\Omega \cdot cm^2$) for each protein layer within a given device configuration from Figure S1, S2. In this study, we used the geometric junction area for area normalization. While this approach serves as a rough approximation for large-area,[14] mechanically contacted junctions, such as Si–Au and Au–EGaIn junctions, it may offer improved accuracy for relatively smaller, precisely controlled junctions, particularly those involving evaporated top electrodes composed of MpD. Additionally, the same methodology was employed to determine the junction resistance of bare devices (electrically shorted configurations without any protein layer) across the Si-Au, Au-EGaIn, and MpD configurations.

**S12. Protein Structural Analysis**

Using the specialized PDB visualization and analysis tools PyMOL and JSmol, we examined the structures of our experimental proteins. For bacteriorhodopsin (bR), we selected the trimeric structure with PDB ID: 1BRR, while for human serum albumin (HSA), we used the monomeric structure with PDB ID: 1BM0.

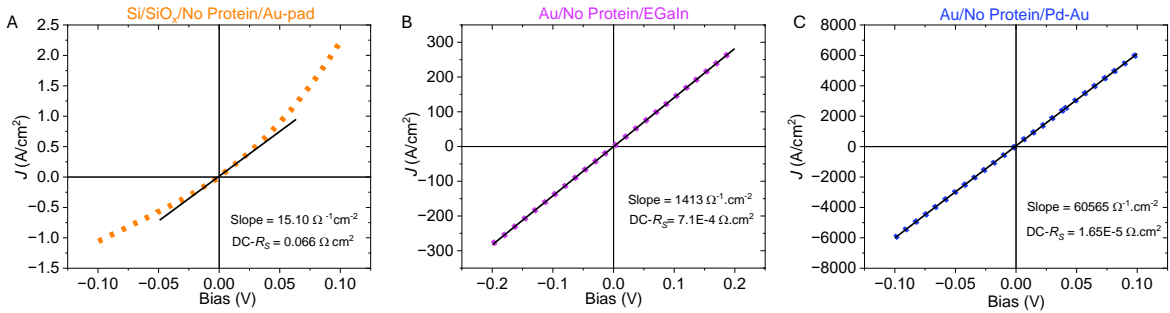

**Figure S1**: (**A-C**) Plots of averaged *J-V* curves for devices without proteins (shorted ones) in each of the three contacting configurations, as indicated in the respective figure legends. The reciprocal of the *J-V* slope yields the *DC* circuit resistance (DC-$R_S$) for each device configuration.

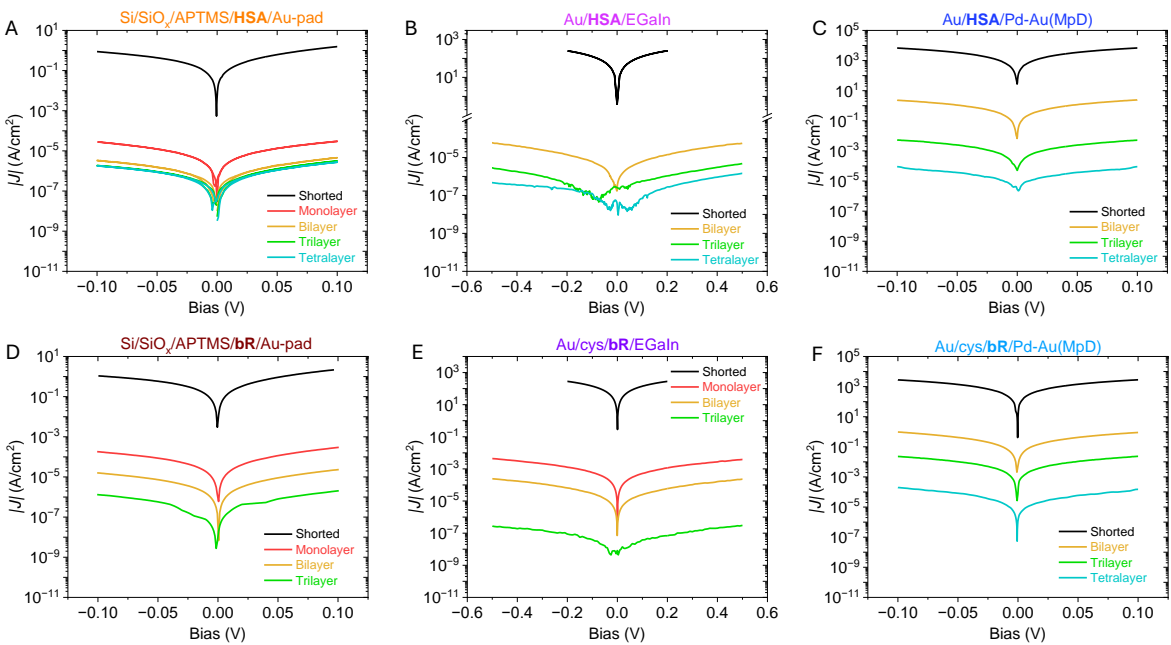

**Figure S2**: Overlay of average *J-V* traces of different protein junctions; the different HSA junctions are shown in the **TOP** row with (**A**) Si-Au, (**B**) Au-EGaIn, and (**C**) Au-Pd (MpD) device configurations. The **BOTTOM** row shows the same, but for bR junctions; in (**D**) Si-Au, (**E**) Au-EGaIn, and (**F**) MpD configuration.

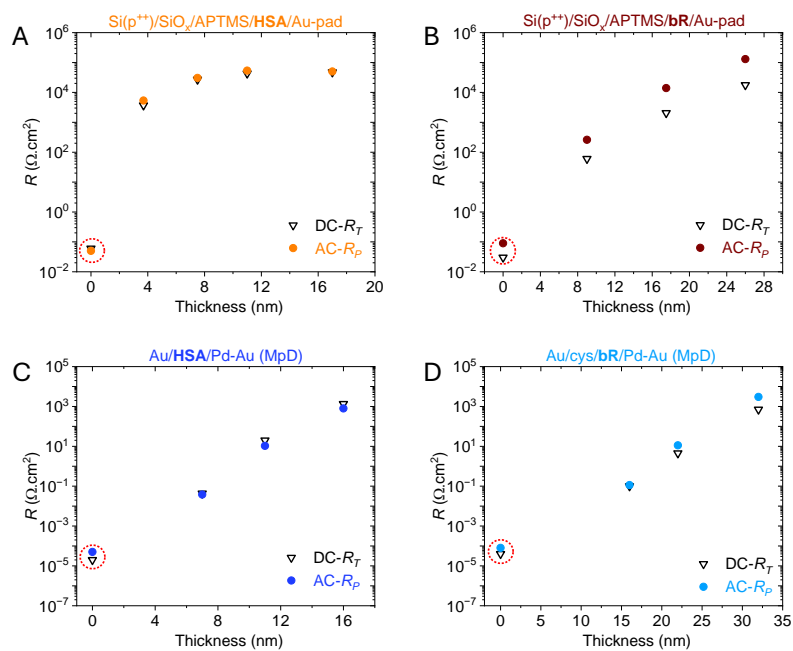

**Figure S3**: Overlay of average protein resistance values, $R_P$, derived from impedance fitting (*AC*) and the total junction resistance, $R_T$, from the direct current (*DC*) measurements, for various HSA and bR junctions. HSA junctions in configurations of (**A**) Si-Au and (**C**) Au-Pd (MpD) and for bR junctions in configurations of (**B**) Si-Au and (**D**) MpD. Red dotted circles show the $R_S$ values of the respective junctions, derived separately from *AC* and *DC* measurements.

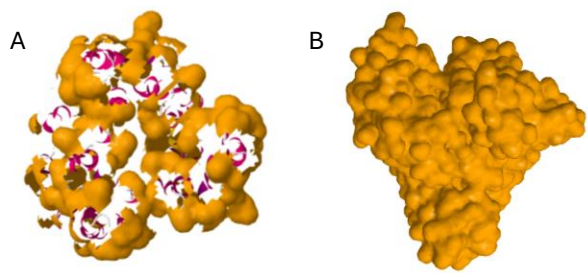

**Figure S4:** Top views of surface-exposed polar/charged residues of **bR** trimer and **HSA** monomer (gold-surface-color) that can interact with the contact surface. The view is chosen so that we look at the protein surfaces that are assumed to bind directly to the substrate, via exposed cysteines for HSA and electrostatically for bR. (**A**) White spaces within the bR trimer structure indicate cavities or hydrophobic regions, while magenta-colored regions highlight visible secondary structures through these cavities. Structures were analyzed using *Jsmol*: (**A**) bR-trimer (PDB: **1BRR**) and (**B**) HSA-monomer (PDB: **1BM0**).

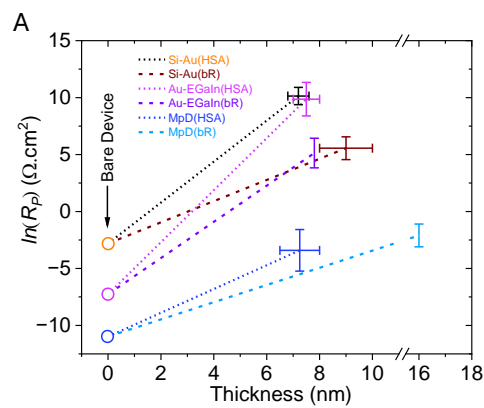

**Figure S5**: $ln(R_P)$ vs. thickness plots for the junctions with mostly comparable protein thicknesses *(d ~8 nm)* with HSA or bR, along with *d* ~16 nm bR in MpD configuration, relative to the resistance of a bare device (*d* = 0).

| **Table S1** | | |
|---|---|---|
| Junctions | Thickness of protein layer (nm) | $\beta$(nm$^{-1}$) |
| Au/**HSA**/Pd-Au | 7.25 ± 0.75 | 1.10±0.10 |
| Au/**HSA**/EGaIn | 7.50 ± 0.50 | 2.30±0.15 |
| Si/SiO$_x$/APTMS/**HSA**/Au-pad | 7.20 ± 0.40 | 1.80±0.10 |
| Au/Cys/**bR**/Pd-Au | 16.0 ± 1.00 | 0.60±0.05 |
| Au/Cys/**bR**/EGaIn | 7.80 ± 0.40 | 1.60±0.05 |
| Si/SiO$_x$/APTMS/**bR**/Au-pad | 9.00 ± 1.00 | 0.95±0.10 |

**Table S1**: Distance decay, *β*, parameter values were obtained from the slope of *ln*($R_P$) vs. thickness (from Figure S5, shown with fitted dotted/dashed lines).

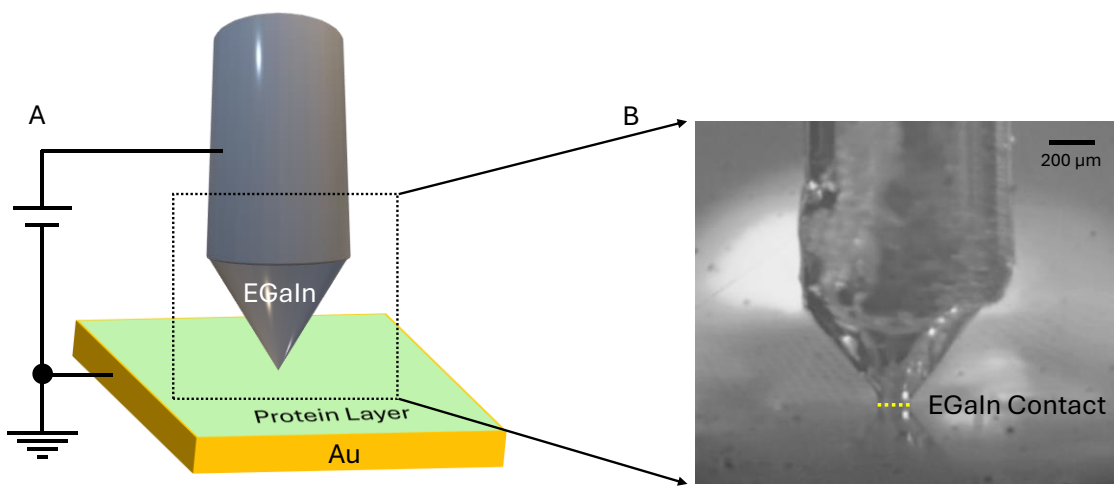

**Figure S6**: (**A**) Schematic representation of the Au/protein/EGaIn-cone (Au-EGaIn) junction used for transport measurements. (**B**) Actual photograph of the junction, showing a side view of the EGaIn-protein contact, with the interface highlighted by a *yellow* dotted line.

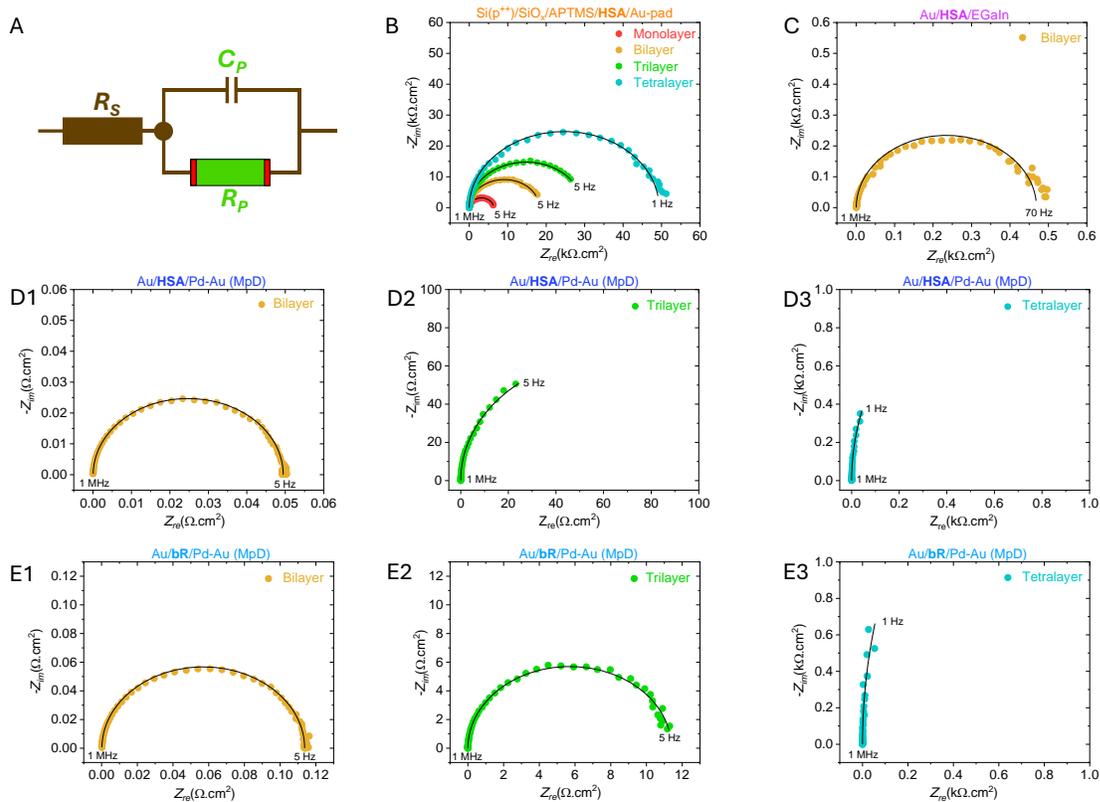

**Figure S7**: **(A)** Schematic representation of the equivalent circuit model used to fit impedance data for protein junctions. **(B-E)** Representative Nyquist plots for various protein junctions; **(B-D)** for **HSA** and **(E)** for **bR**, **(B)** in Si-Au configuration with different layers as indicated in the figure legends, **(C)** of HSA bilayer in Au-EGaIn configuration, **(D)** and **(E)** for HSA and bR in the MpD configuration; **(D1, E1)** a bilayer, **(D2, E2)** a trilayer, and **(D3, E3)** a tetralayer.

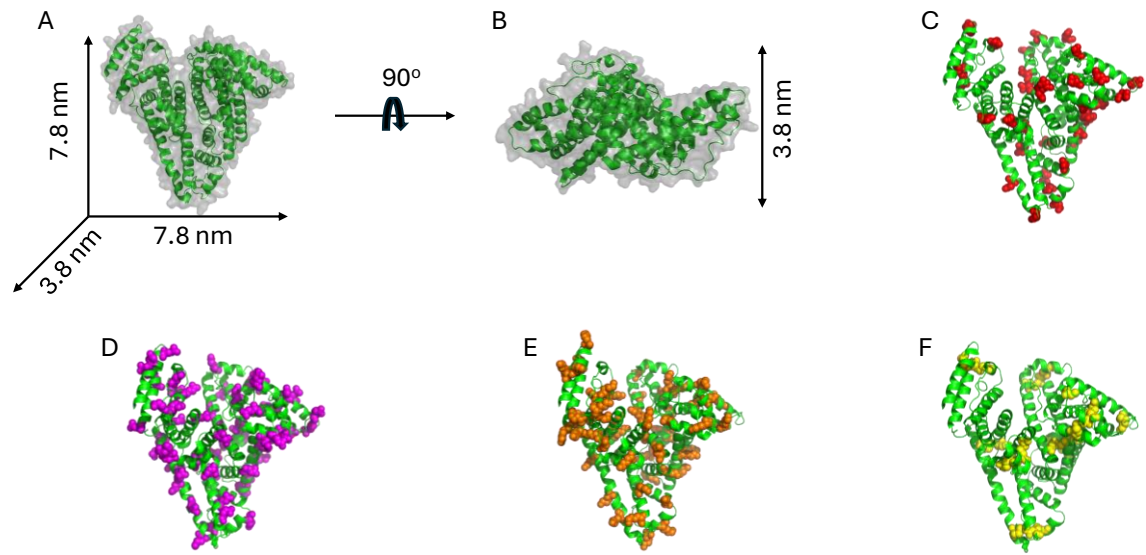

**Figure S8**: The crystal structure of human serum albumin (HSA), analyzed using PyMOL, based on the Protein Data Bank, PDB, entry 1BM0 (monomer). (**A**) A quasi 3-D representation of the HSA monomer; the green shows the protein's secondary and tertiary structure; the outer-most part of the semi-transparent gray volume defines the protein's outer boundary; the three principal dimensions are indicated by arrows. (**B**) The preferred surface orientation of the HSA monomer is the 'lying down' configuration. (**C**) - (**F**): Specific amino acid residues with potential activity are highlighted on the HSA structure; how many of each are there appears in parentheses: (**C**) Aspartic acid residues (**35**) in red, (**D**) Glutamic acid residues (**62**) in magenta, (**E**) Lysine residues (**58**) in orange and (**F**) Cysteine residues (**35**) in yellow.

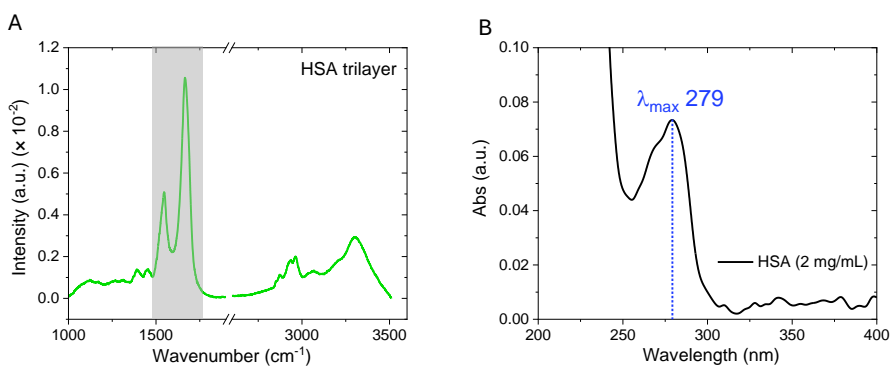

**Figure S9**: (A) PM-IRRAS spectral signature of HSA trilayer (~11.5 nm) on Au-substrate over a full spectral range showing the typical amide I and II characteristics (shaded zone) as seen also for other solid state protein layers (*see Section S7 in SI*). (B) A typical UV-Vis spectra of 2 mg/mL HSA solution in 20 mM phosphate buffer at 5.5 pH in 150 mM NaCl medium, collected from Nanodrop with 1 mm path length.

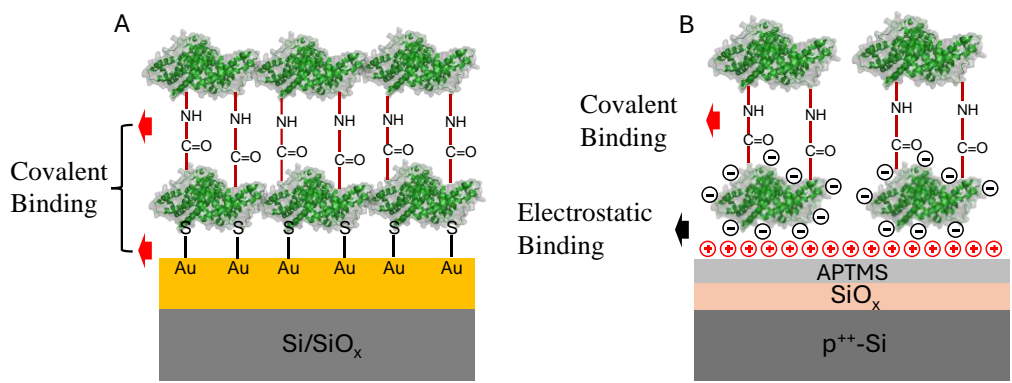

**Figure S10**: Schematic representation of layer-by-layer HSA multilayer on (A) Au-substrate/MpD and (B) $p^{++}$-Si/SiO$_x$/APTMS, with the interface charges indicated. Net surface charges from carboxylate groups (aspartic and glutamic acids) on the protein envelope are shown for the first layer.

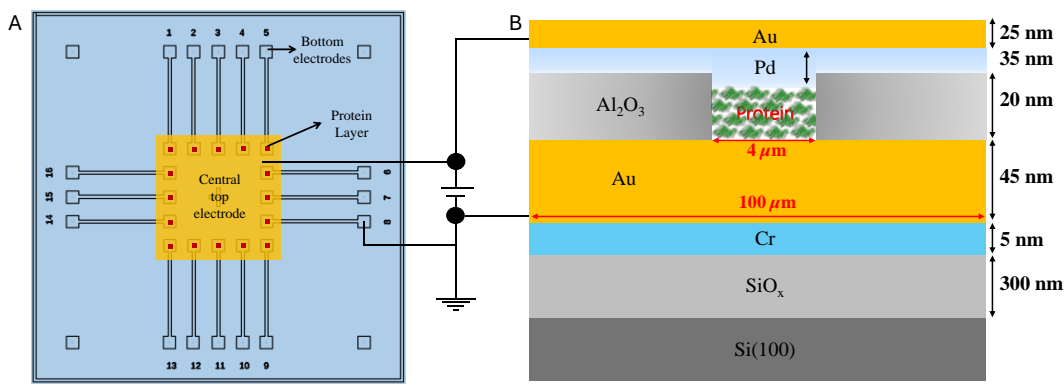

**Figure S11**: (A) Schematic representation of an MpD chip shows an array of 16 devices used for the measurements. The top electrode, a Pd-Au layer (yellow square), serves as the central common electrode, and separate bottom electrodes are used for biasing each device. The red squares highlight the protein layers within the micropores. (B) A cross-sectional schematic view of the complete MpD device, showing the stacked protein layers. The thicknesses (in nanometers) of all the device layers are marked by vertical black arrows. Additionally, a side view of the protein-covered part of the individual bottom electrodes (4 × 4 $\mu m^2$) is provided at the center of the micropore area (100 × 100 μm2), as indicated by the red horizontal arrow in one of the two lateral directions.

| Table S2A: HSA Layer Thickness (nm) from Ellipsometry | | |
|---|---|---|
| HSA layers | Si/SiO$_x$/APTMS/**HSA** | Au/**HSA** |
| Monolayer | 3.4-4.0 | 2.8-3.5 |
| Bilayer | 6.8-7.6 | 6.5-8.0 |
| Trilayer | 10.5-11.5 | 10.8-12.0 |
| Tetralayer | 15.5-17.0 | 15.8-16.4 |

| Table S2B: bR Layer Thickness (nm) from Ellipsometry | | |
|---|---|---|
| bR layers | Si/SiO$_x$/APTMS/**bR** | Au/cys/**bR** |
| Monolayer | 8.2-9.4 | 7.6-8.1 |
| Bilayer | 15-17 | 15-17 |
| Trilayer | 24-30 | 22-27 |
| Tetralayer | -- | 32-35 |

**Table S2**: (**S2A**) Ellipsometry-derived thicknesses for HSA and (**S2B**) bR layers on both Si and Au substrates. The average thicknesses of the linker layers were: (APTMS) 0.5±0.1 nm, cysteamine (cys) 0.6±0.1 nm, and the SiO$_X$ layer was 0.9±0.2 nm thick.

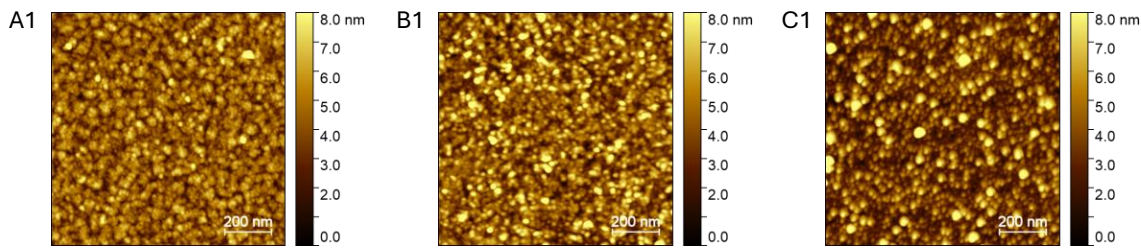

**Figure S12**: Tapping mode AFM topography images of HSA monolayers on different electrode (substrate) surfaces: (A) Au substrate, (B) micropore region of MpD, and (C) $p^{++}$-Si/SiOx/APTMS. The corresponding RMS roughness values are $1.0 \pm 0.05$ nm, $1.2 \pm 0.2$ nm, and $1.2 \pm 0.1$ nm, respectively.

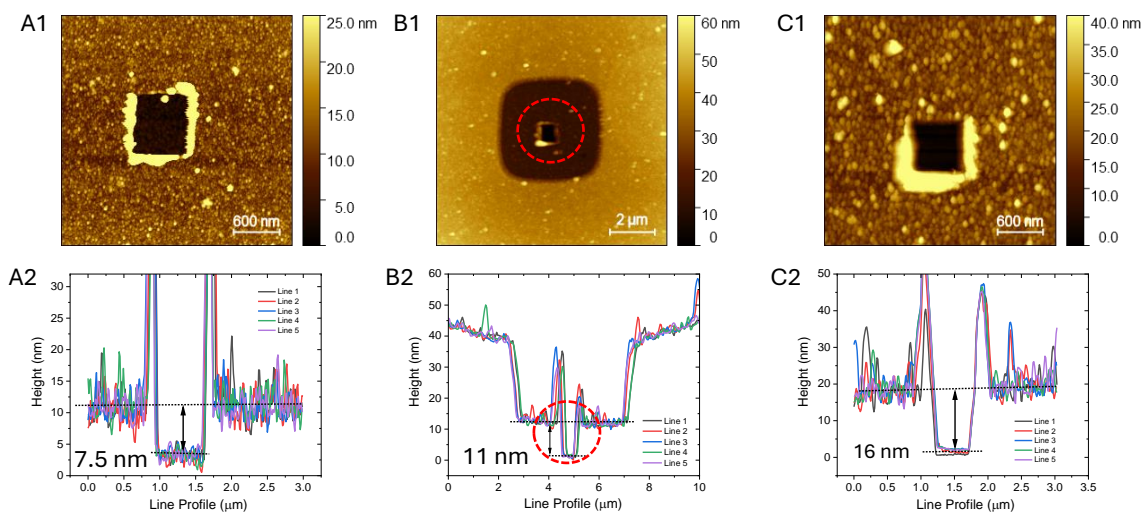

**Figure S13**: **Left column**: HSA bilayer on an Au substrate; **Middle column**: HSA trilayer in MpD; **Right column**: HSA tetralayer on Si/SiOx/APTMS. Panels (A1, B1, C1) show AFM topography images of scratched regions (marked by black squares), while panels (A2, B2, C2) present cumulative five-line profiles across the corresponding scratched areas. From these profiles, we estimate the thickness of the HSA multilayers: **7.5 ± 0.5 nm** for the bilayer, **11 ± 0.5 nm** for the trilayer, and **16 ± 1.0 nm** for the tetralayer. In panel (B1), the scratched region (highlighted with a dotted red circle) is located within the center of the micropore area (~4 × 4 μm²) of the MpD surface.

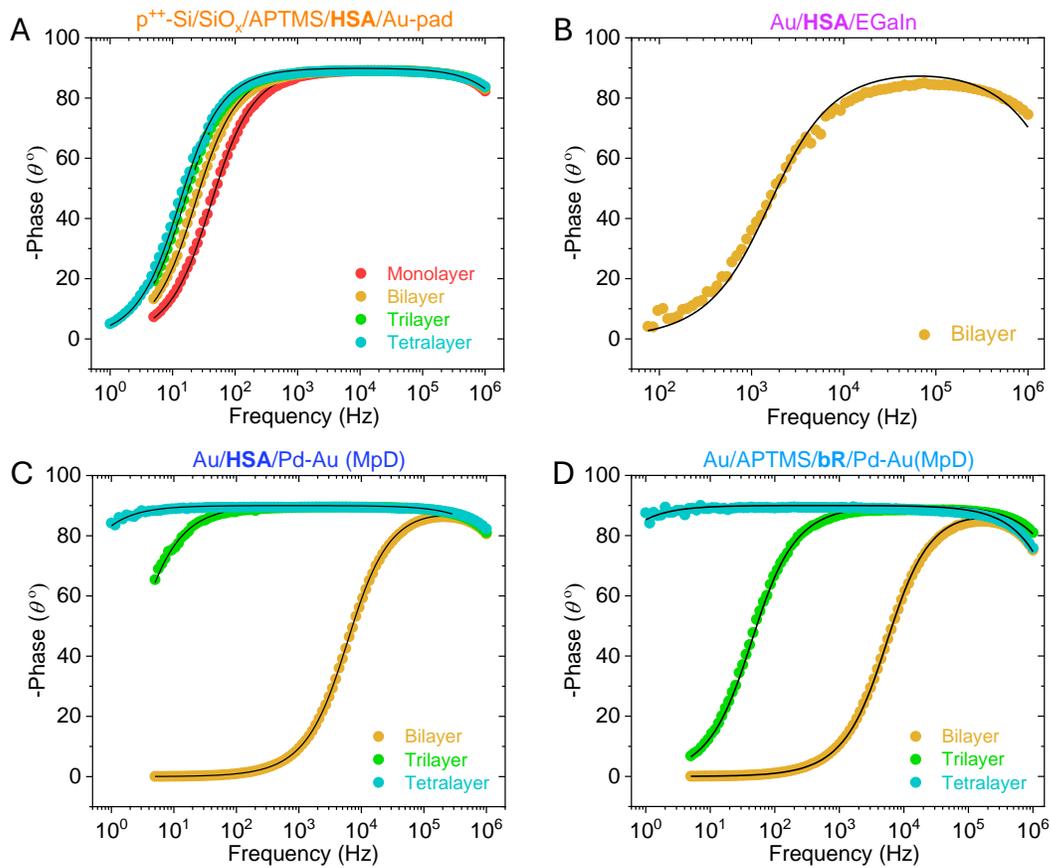

**Figure S14**: Bode plots (phase-shift vs. operating frequency) for protein junctions in the different device configurations. (**A-C**) HSA: (**A**) Si-Au, (**B**) Au-EGaIn, and (**C**) MpD. (**D**) is for the bR junctions in MpD configuration. In ref.[1] the corresponding results for the Si-Au junction are given. The colored dotted line represents the experimental data, while the solid black line shows the impedance-fitted data based on the equivalent circuit model presented in Figure S7A.

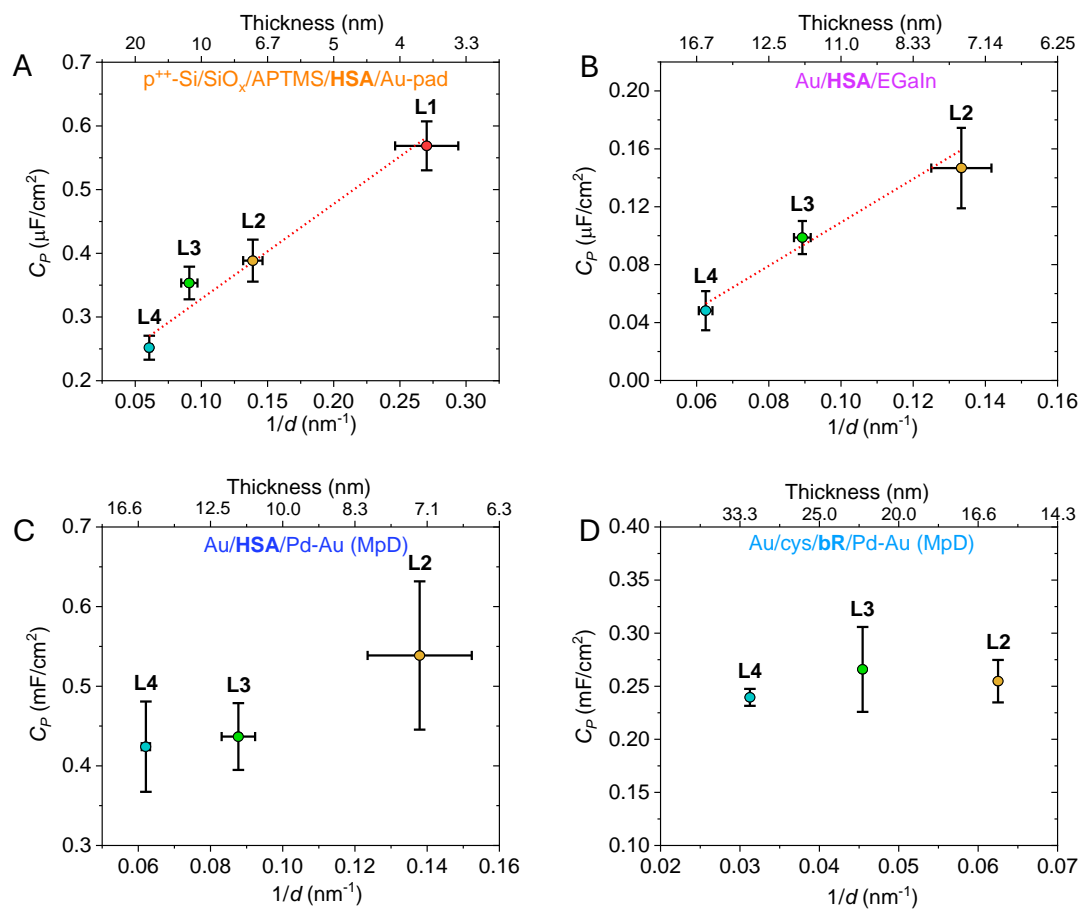

**Figure S15: (A-C):** Plots of $C_P$ vs. $1/d$ of different HSA junctions in **(A)** Si-Au, **(B)** Au-EGaIn, and **(C)** Au-Pd (MpD) configurations and **(D)** bR junctions in the MpD configuration. The different protein layers are labeled as L1, L2, L3, and L4, representing a monolayer, bilayer, trilayer, and tetralayer, respectively.